\documentclass[aps,prl,showpacs,twocolumn,superscriptaddress,nobibnotes]{revtex4-1}

\RequirePackage{subfigure}
\usepackage{lipsum}
\usepackage{graphicx}
\usepackage{color}
\usepackage{enumerate}
\usepackage{amsmath}
\usepackage{bm}
\usepackage{diagbox}
\usepackage{amssymb}
\usepackage{stmaryrd}
\usepackage{multirow}
\usepackage[normalem]{ulem}
\usepackage{float}
\usepackage{longtable,booktabs}
\usepackage[dvipsnames,table,xcdraw]{xcolor}
\usepackage[normalem]{ulem}
\usepackage{physics}
\usepackage{url}
\usepackage{soul} 
\usepackage[colorinlistoftodos]{todonotes}
\usepackage{verbatim}
\usepackage{graphicx}
\usepackage{appendix}
\definecolor{darkblue}{rgb}{0.1,0.2,0.6} \definecolor{darkred}{rgb}{0.8,0.1,0.2}
\usepackage[colorlinks,citecolor=darkblue,linkcolor=darkred,urlcolor=darkblue]{hyperref}
\usepackage[all]{hypcap} 
\usepackage[draft]{pdfcomment}

\def\beq{\begin{equation}}
\def\eeq{\end{equation}}

\begin{document}
\long\def\/*#1*/{}

\title{Backflow Transformations via Neural Networks \\ for Quantum Many-Body Wave-Functions}

\author{Di Luo}
\affiliation{Institute for Condensed Matter Theory and Department of Physics, University of Illinois at Urbana-Champaign, IL 61801, USA} 

\author{Bryan K.  Clark}
\affiliation{Institute for Condensed Matter Theory and Department of Physics, University of Illinois at Urbana-Champaign, IL 61801, USA} 

\begin{abstract}

Obtaining an accurate ground state wave function is one of the great challenges in the quantum many-body problem.  In this paper, we propose a new class of wave functions, neural network backflow (NNB).  The backflow approach, pioneered originally by Feynman \cite{Feynman_1956prl_He4}, adds correlation to a mean-field ground state by transforming the single-particle orbitals in a configuration-dependent way.  NNB uses a feed-forward neural network to learn the optimal transformation via variational Monte Carlo.  NNB directly dresses a mean-field state, can be systematically improved and directly alters the sign structure of the wave-function.  It generalizes the standard backflow\cite{Sorella_2008prb_backflow} which we show how to explicitly represent as a NNB.  We benchmark the NNB on Hubbard models at intermediate doping finding that it significantly decreases the relative error, restores the symmetry of both observables and single-particle orbitals, and decreases the double-occupancy density. Finally, we illustrate interesting patterns in the weights and bias of the optimized neural network.

\end{abstract}
 
\maketitle

\textit{Introduction.}---A key question in strongly correlated quantum systems is to obtain an approximation for the ground state wave function. This is especially important for Fermion systems in two or more dimensions where only approximate or exponentially costly methods for evaluating observables of quantum systems exist.  Early attempts for writing down variational Fermion wave-functions, such as Slater determinants \cite{Slater_1929_SlaterDet} and BCS wave-functions\cite{Bardeen_1957PRL}, focused on finding the ground state of a mean field Hamiltonian which best matched the interacting ground state.  Since these early attempts more sophisticated wave-functions have been developed which dress these mean-field starting points including Slater-Jastrow \cite{Jastrow_1955PRL_Jastrow,Martin_1994PRL_Jastrow}, Slater-Jastrow-Backflow \cite{Feynman_1956prl_He4,Ceperley_1993prb_backflow} and iterative backflow \cite{Taddei_2015prb_iterative} which has recently been described as a non-linear network \cite{Ruggeri_2017PRL_nonlinear}.  These wave-functions have the advantage that the mean-field starting point can directly incorporate the basic physics of the problem.  

Instead of starting from a dressed mean-field, many other classes of wave-functions are parameterized by a tuning parameter $D$ which interpolates from a trivial state at small $D$ to a universal wave-function spanning the entire Hilbert space at exponential $D$. Examples of such wave-functions include matrix-product states\cite{white_1992PRL_dmrg,vidal_2003PRL_mps},  other forms of tensor networks\cite{vidal_2007PRL_MERA,Verstraete_2006PRL_PEPS,vidal_2003PRL_mps}, Huse-Elser states\cite{Huse_1988PRL_Huse-Elser,Mezzacapo_2009_entangled-plaquette,Marti_2010_graph_tensornet}, and string-bond states \cite{Ignacio_2008PRL_stringbond}.
Recently, wave-functions based on neural network primitives, such as restricted Boltzmann machines (RBM) and feed forward neural network(FNN), have been introduced with similar universal properties\cite{Carleo_2017sci_rbm,Gao_2017nature,DengDongLing_2017PRX,Glasser_2018prx_nnstate,Chen_2018prb_equivalence,Carleo_2017sci_rbm,Yusuke_2017prb_rbm,Carleo_2018arxiv_drmb,Freitas_2018arxiv_drmb,Cai_2018prb_ann,Carleo_2018_RBM_excited_state,weinan_2018_DNN_schrodinger,Carleo_2018npj_VAE,Saito_2018_ANN_quantum_few_body,Saito_2018JPS_Boson_lattice,Saito_2018_ANN_quantum_few_body,Moore_2017_NN_tensornet,Dongling_2017PRB_ml_topo,WangLei_2018PRB_RBM_tensornet,Clark_2018JPA_NQS,Carleo_2017_QST,Kaubruegger_2018PRB_Chiral,Liang_2018_CQNS,Carleo_2018_quantum_computation,Kaubruegger_2018_transfer_matrix_ANN,Lu_2018_topo_RBM,Dima_2018_AI}:  as the number of hidden neurons increases, the neural network state can represent all probability distribution although may require complex weights to represent the sign structure of Fermion wave-functions.  A recent attempt to incorporate RBM into Fermion states by using the RBM as a more general Jastrow \cite{Yusuke_2017prb_rbm} shows promise but was still restricted to the sign-structure of the underlying mean-field ansatz. Even though general neural networks could alter the sign-structure, it may struggle with capturing the underlying mean-field physics both in terms of the number of neurons required as well as optimization. 

\begin{figure}[H]
  \centering
  \includegraphics[scale=0.35]{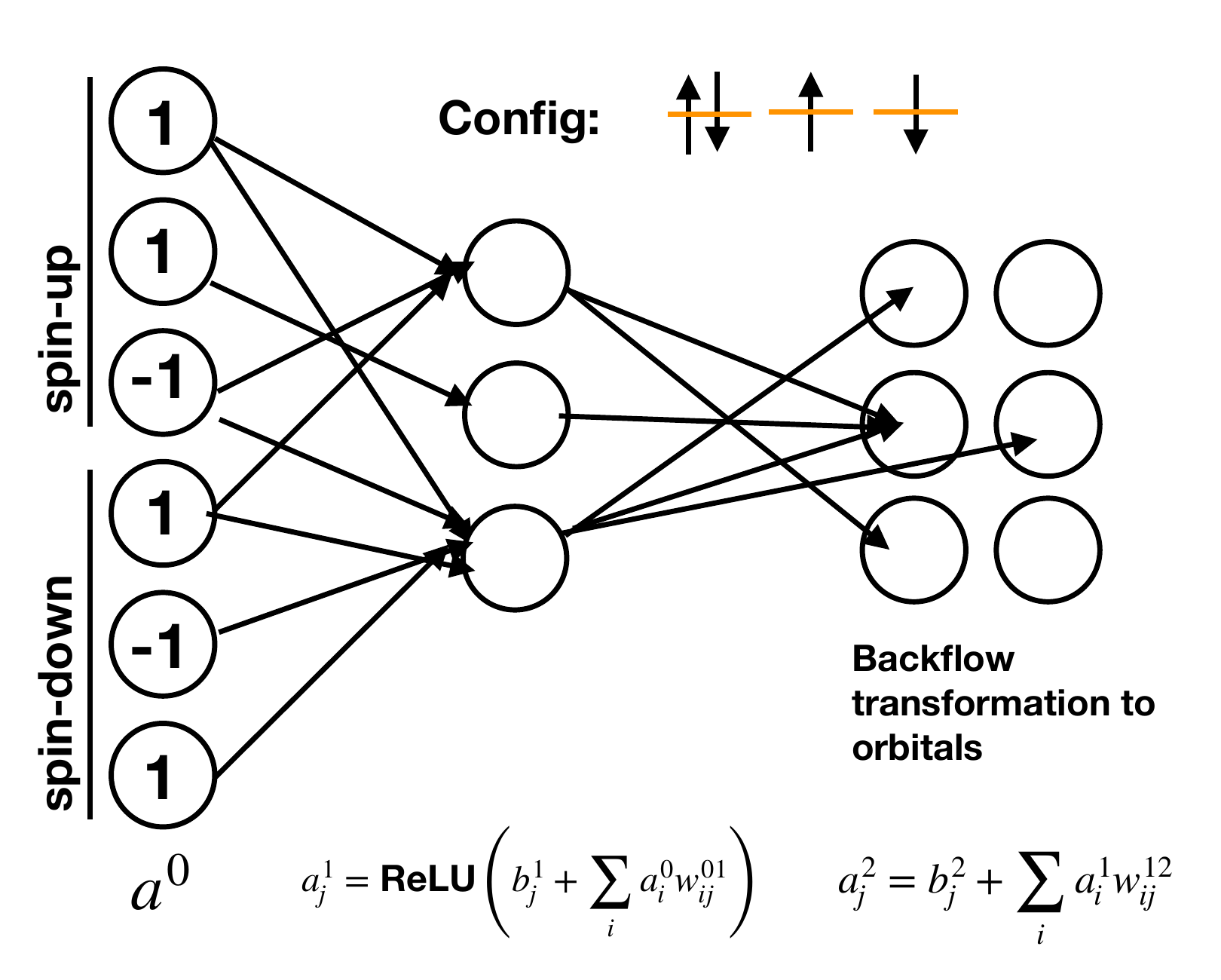}
  \caption{Cartoon of (spin-up) neural network being used in this work for $a^{NN}$ (other transformations are similar) with input shown for the configuration displayed of 4 electrons on 3 sites.  Every layer is fully connected with arrows but only a fraction of them are shown for image clarity.  The input layer is set by the configuration.  Parameters $b_j$ and $w_{ij}$ are bias and weights to be optimized. The output layer is a $n_\textrm{up-electrons} \times n_\textrm{sites}$ (i.e $2 \times 3$) matrix which will be the backflow transformation added to the single particle orbitals. }\label{fig:BFNN}
\end{figure}

In this work, we propose a new class of wave-functions, the Neural Network Backflow (NNB), which dresses a  mean-field wave-function, can make changes to the sign structure directly, and can be systematically improved by increasing the number of hidden neurons and is able to be made theoretically exact in the limit of enough neurons.  To accomplish this we use a feed-forward neural network (FNN), not in the standard approach of returning a wave-function amplitude, but instead to transform the single particle orbitals in a configuration dependent way; these orbitals are then used in the mean-field wave-function.  Wave-functions with configuration-dependent orbitals are known as a backflow wave-function\cite{Feynman_1956prl_He4,Ceperley_2003pre_backflow,Tocchio_2011prb_backflow,Tocchio_2009jop_backflow,Carleo_2011prb_backflow,Sorella_2009prb_backflow,Sorella_2008prb_backflow,Drummond_2006chp_backflow,Ceperley_1993prb_backflow,Taddei_2015prb_iterative,Ruggeri_2017PRL_nonlinear,Tocchio_2010PRB_Interaction,Ceperley_1998PRB_backflow,Drummond_2006PRE_backflow,Lee_1981PRL_He3,Lee_1981PRL_Fermion_fluid,Carlson_1989PRL_FMC,Ceperley_2006PRB_backflow}.

\textit{Background.}---Mean field theory - approximating the ground state of a quartic Hamiltonian by the ground state, $\psi_{MF}$, of a quadratic Hamiltonian - is a powerful first step to understanding correlated quantum systems.    Various mean fields lead to  different types of ground states including Slater determinants, 
\begin{align}
\label{Eq:SD}
\psi_{SD}(\bm{r})  =  & \det[M^{SD,\uparrow}] \det[M^{SD,\downarrow}]; \\
M_{ik}^{SD,\sigma} = & \phi_{k\sigma}(r_{i\sigma})
\end{align}
and Bogoliubov de Gennes wave-functions, 
\begin{align}
\label{Eq:BDG}
\psi_{BDG}(\bm{r})= & \det[\Phi] \\ 
\Phi_{ij} = &  \sum^{N}_{k,l=1} \phi_{k\uparrow}(r_{i,\uparrow})\hspace{0.1cm}S_{kl} \hspace{0.1cm} \phi_{l\downarrow}(r_{j,\downarrow}) 
\end{align}
where $\phi_{k\sigma}$ is the $k$'th single particle orbital and $r_{i,\sigma}$ is the position of the $i$'th particle of spin $\sigma$. Eq.~\eqref{Eq:SD} only takes the occupied orbitals while Eq.~\eqref{Eq:BDG} is summed over both occupied and unoccupied orbitals.

Mean field states are uncorrelated by construction.  The simplest way to capture correlation physics is through the introduction of a Jastrow giving $\psi_\textrm{Jastrow}(\bm{r}) =  \exp[-U(\bm{r})]\psi_{MF}(\bm{r})$, where $U(\bm{r})$ is an arbitrary function. In this work, we always use a charge Jastrow $U(\bm{r})=\frac{1}{2} \sum_{i,j} v_{ij} n_i n_j$, where $n_i$ is the charge density, $v_{ij}$ is the variational parameters. While Jastrow factors can introduce many-body correlations, they can't modify the mean-field's sign structure. One approach to add additional sign-structure modifying correlation  is through a backflow correction \cite{Ceperley_2003pre_backflow,Tocchio_2011prb_backflow,Tocchio_2009jop_backflow,Carleo_2011prb_backflow,Sorella_2009prb_backflow,Sorella_2008prb_backflow,Drummond_2006chp_backflow,Ceperley_1993prb_backflow,Taddei_2015prb_iterative,Ruggeri_2017PRL_nonlinear,Tocchio_2010PRB_Interaction,Ceperley_1998PRB_backflow,Drummond_2006PRE_backflow,Lee_1981PRL_He3,Lee_1981PRL_Fermion_fluid,Carlson_1989PRL_FMC,Ceperley_2006PRB_backflow}  which introduces correlations by having the single-particle orbitals act on a configuration-dependent quasi-particle position.  
On the lattice, the backflow approach instead uses a configuration-dependent mean-field \cite{Sorella_2008prb_backflow,Sorella_2009prb_backflow,Tocchio_2009jop_backflow,Tocchio_2011prb_backflow} - i.e. the quadratic Hamiltonian or single-particle orbitals  $\phi_{k,\sigma}^b(r_i;\bf{r})$ depend not only on the position $r_i$ but on all other electron positions $\bf{r}$; our NNB builds on top of this formulation of backflow.

\textit{Neural Network Backflow.}---The NNB uses a FNN to modify the single particle orbitals for a spin $\sigma$,
\begin{equation}
\phi_{k\sigma}^{b}(r_{i,\sigma};\bm{r})= \phi_{k\sigma}(r_{i,\sigma}) +  a^{NN}_{ki,\sigma}(\bm{r}) \label{eq:BFNN2}
\end{equation}
where each value of  $a^{NN}_{ij,\sigma}$ is represented by an output neuron of the FNN.  We use one neural net for each of $\sigma \in \{\uparrow, \downarrow \}$. This is to be contrasted with the standard backflow\cite{Tocchio_2011prb_backflow} parameterization, 
\begin{eqnarray}
\phi_{k\sigma}^{b}(r_{i,\sigma};\bm{r}) & = & \phi_{k\sigma} + \sum_{j} \eta_{ij,\sigma}  \phi_{k\sigma}(r_{j,\sigma}) \nonumber \\ 
\eta_{ij,\sigma} &  = & t D_i H_j \theta_{|i-j|,\sigma} \label{eq:eta_ij}
\end{eqnarray}
with $D_i = n_{i,\uparrow}n_{i,\downarrow}$, $H_i=(1-n_{i,\uparrow})(1-n_{i,\downarrow})$. $\theta_{1,\sigma}$ and $\theta_{2,\sigma}$ are the only non-zero variational parameters.

Interestingly, the backflow transformation of Eq.~\eqref{eq:eta_ij}, can be represented as a neural network for $a^{NN}_{ij,\sigma}(\bm{r})$ with three hidden layers and a linear number of neurons; an explicit construction will be given in the next section. This ensures that there exists a three layer neural network which is at least as good as the standard backflow transformation.

\begin{figure}[H]
  \centering
      \includegraphics[scale=0.2]{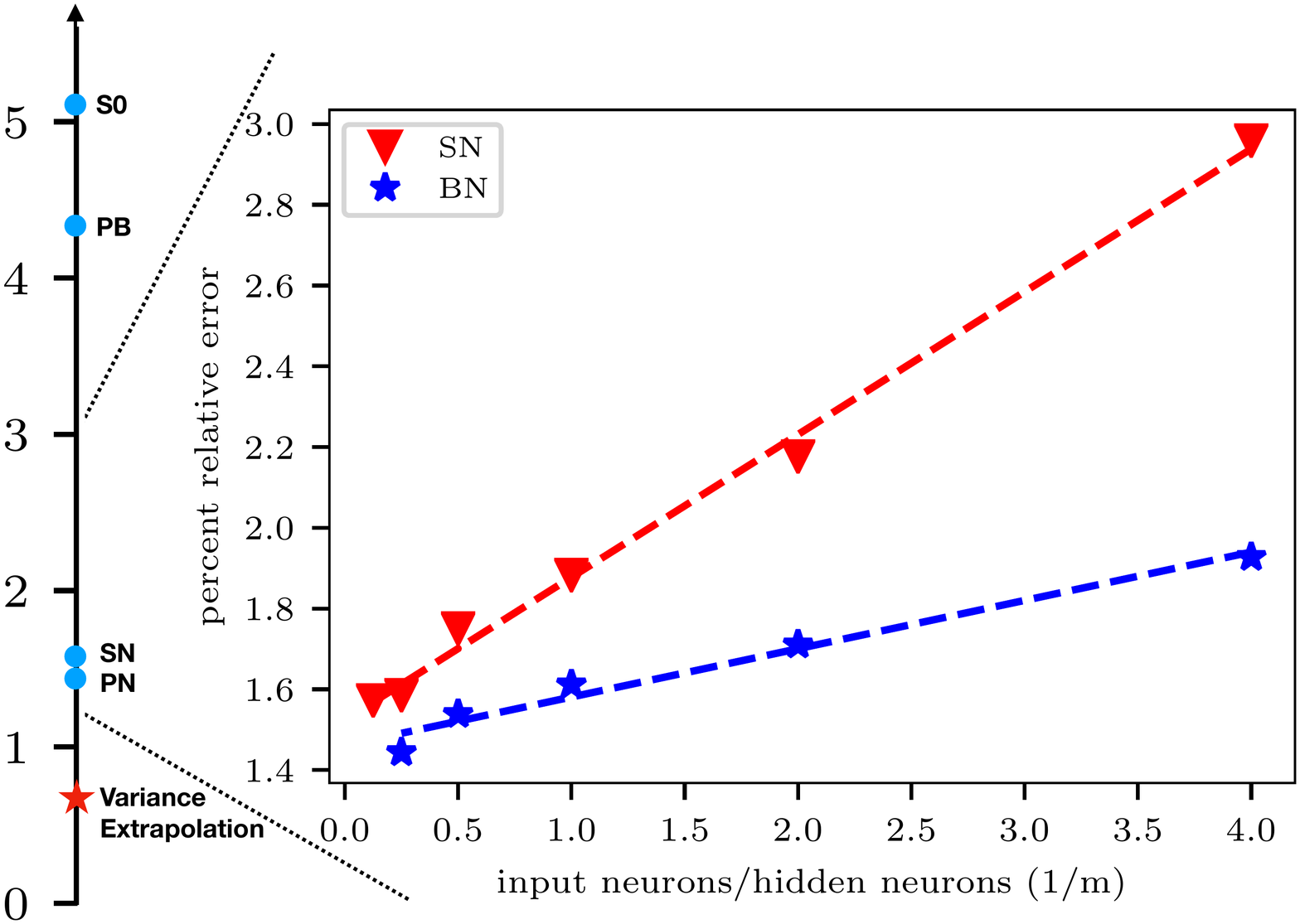}
       \includegraphics[scale=0.7]{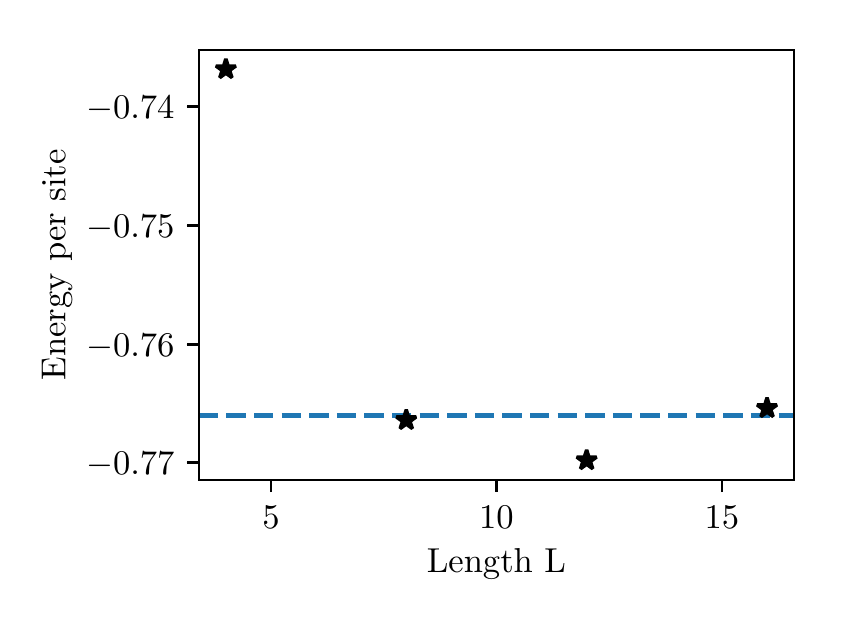}
  \caption{\textbf{Top Left:} Percentage relative error from the exact ground-state energy of Eq.~\ref{eq:Hubbard} ($E=-11.868$\cite{Dagotto_1992prb_ED}) on $4 \times 4$ Hubbard model at U/t=8, n=0.875, for various classes of wave-functions. The star is the variance extrapolation result of $\Psi_\textrm{PN}$ (see Supplementary Material  \cite{supp} Sec. I). \textbf{Top Right:} Percentage relative energy error ,($E_{exact}$-$E_{NNB}$)/$E_{exact}$*100 $\%$, as a function of $1/m$ for NNB.  Statistical error bars are shown but smaller than the marker size.  \textbf{Bottom:} Variance extrapolated energy per site for Hubbard model with U/t=8, n=0.875 with system size $L \times 4$ for $L=4,8,12,16$. The dash line is the DMRG energy per site ($-0.7659 \pm 4 \times 10^{-5}$) for system size $\infty \times 4$ (PBC, open) \cite{Hubbard_stripe}.
  }\label{fig:E_min}
\end{figure}

\begin{table*}[t]
  \centering
  \begin{tabular}{|c|c|c|c|} \hline
  method & backflow transformation & mean field  & variational functions \\\hline
  $\psi_\textrm{S0}$ & NA. & Eq.~\eqref{Eq:SD} & $\phi_{k\uparrow}(r_{i,\uparrow})$,$\phi_{k\downarrow}(r_{i,\downarrow})$,$v_{ij}$ \\\hline
  $\Psi_\textrm{SN}$ & $\phi_{k\sigma}^{b}(r_{i,\sigma};\bm{r})= \phi_{k\sigma}(r_{i,\sigma}) +  a^{NN}_{ki,\sigma}(\bm{r})$ & Eq.~\eqref{Eq:SD} & $a^{NN}_{ki,\sigma}(\bm{r})$,$v_{ij}$ \\\hline
  $\Psi_\textrm{PB}$ & $\phi_{k\sigma}^{b}(r_{i,\sigma};\bm{r}) =  \phi_{k\sigma}(r_{i,\sigma}) + \theta_{1\sigma}  \sum_{j} t D_i H_j \phi_{k\sigma}(r_{j,\sigma}) + \theta_{2\sigma}  \sum_{j} t D_i H_j \phi_{k\sigma}(r_{j,\sigma})$ & Eq.~\eqref{Eq:BDG} & $\theta_{1\sigma}$,$\theta_{2\sigma}$,$S_{kl}$,$v_{ij}$ \\\hline
  $\Psi_\textrm{PN}$ & $\phi_{k\sigma}^{b}(r_{i,\sigma};\bm{r})= \phi_{k\sigma}(r_{i,\sigma}) +  a^{NN}_{ki,\sigma}(\bm{r})$; $S_{kl}(\bm{r})=S_{kl}+d^{NN}_{kl}(\bm{r})$ & Eq.~\eqref{Eq:BDG}  & $a^{NN}_{ki,\sigma}(\bm{r})$,$d^{NN}_{kl}(\bm{r})$,$v_{ij}$ \\\hline
  \end{tabular}
  \caption{Wave Function Ansatzs}
  \label{tb:ansatz}
\end{table*}
We consider two NNB wave-functions, $\Psi_{SN}$ and $\Psi_{PN}$, implemented on top of a Slater Determinant and BCS pairing wave-functions respectively.  The neural nets used in these wave-functions are similar although $\Psi_{SN}$ has outputs which only correspond to the occupied orbitals, while the outputs of $\Psi_{PN}$ correspond to all the orbitals. In addition, for $\Psi_{SN}$ there are only two neural nets (one for each of the spin-up and spin-down orbitals) while for $\Psi_{PN}$ there is an additional neural net used to generate a system dependent $S_{kl}(\bm{r})$. This is implemented by letting $S_{kl}(\bm{r})=S_{kl}+d^{NN}_{kl}(\bm{r})$, where $d^{NN}_{kl}(\bm{r})$ is represented by an FNN (in this work always fixed to 16 hidden neurons) that inputs the system configuration $\bm{r}$ and outputs the symmetric matrix correction $d^{NN}_{kl}$. Notice that $\Psi_\textrm{PN}$ is trivially a superset of $\Psi_{SN}$. 

Although various architectures can be used, we adopt a three-layer fully-connected FNN for each of the functions $a^{NN}_{ki,\sigma}$ and $d^{NN}_{kl}$ (see Fig.~\ref{fig:BFNN}).   The input layer has $2N$ neurons with neuron $i$ (neuron $i+N$) outputting 1 if there is spin up (spin down) on site $i$ and -1 otherwise, where $N$ is the total system size. The hidden layer contains $mN$ hidden neurons for constant $m$ with Rectifier Linear Units (ReLU) \cite{Xu_2015arXiv_relu} activation functions.   The output layer then contains $O(N^2)$   neurons specifying the values of the respective functions.  Gradients are computed in the standard way using variational Monte Carlo (see Supplementary Material  \cite{supp} Sec. II) which requires evaluating the derivative of the wave-function with respect to the weights and bias in the neural network.  Derivatives for FNN are typically taken using back-propagation. Because the wave-function is a determinant of a matrix generated by the neural-network output, we evaluate this full derivative by envisioning this determinant as an additional final layer of the neural network and then performing back-propagation including this layer. This ensures the cost of computing all the derivatives is of the same order as the evaluation of the wave-function (see Supplementary Material  \cite{supp} Sec. III).   Optimization is performed by stepping each parameter in the direction of the gradient with a random magnitude, which helps us avoid shallow local minima \cite{sandvik_2007_SR}, or by the RMSPROP method \cite{opt}. 

\begin{figure}[H]
  \centering
 \includegraphics[scale=0.09]{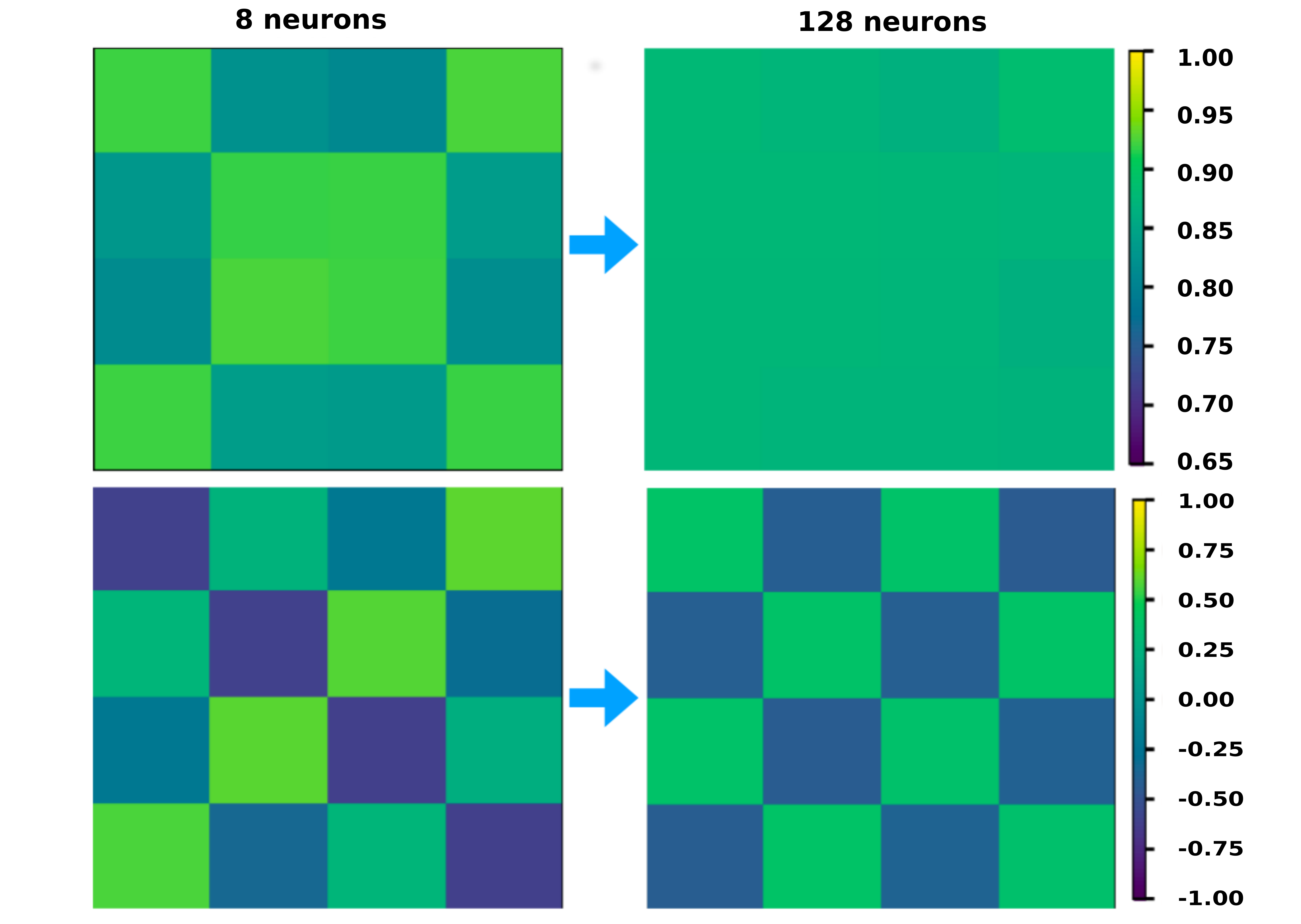}
  \caption{Charge Density (Top) and Spin Density (Bottom) from $\Psi_\textrm{PN}$ with  8 hidden neurons (left) and  128 hidden neurons (right) on $4 \times 4$ Hubbard model at U/t=8, n=0.875. 
  }\label{fig:Density1}
\end{figure}

The computational complexity of the NNB implemented with a single layer of $O(mN)$ hidden neurons scales as $O(mN^4)$ per sweep (i.e. after $N$ electrons move) for forward and backward propagation and $O(N^4)$ per sweep for the evaluation of the mean-field determinant.  This is similar to the scaling of standard backflow. 

\textit{Explicit construction of standard backflow.}---In this section, we provide an explicit construction which represents the standard backflow transformation in the form of Eq.~\eqref{eq:eta_ij} written as a NNB. 

In Eq.~\eqref{eq:eta_ij},  $\eta_{ij,\sigma} = t D_i H_j \theta_{|i-j|,\sigma}=t n_{i,\uparrow}n_{i,\downarrow} h_{j,\uparrow}h_{j,\downarrow}  \theta_{|i-j|,\sigma}$, where $h_{j,\sigma}=1-n_{j,\sigma}$, $\theta_{1,\sigma}$ and $\theta_{2,\sigma}$ are the only non-zero variational parameters. We first demonstrate that $\eta_{ij,\sigma}$ can be presented by a two layer neural network with input layer as $(\sigma_{1},...,\sigma_{N},\sigma_{N+1},...,\sigma_{2N})$ where $\sigma_{i}=2n_{i,\uparrow}-1$ and $\sigma_{i+N}=2n_{i,\downarrow}-1$. By construction, $n_{i,\sigma}$ and $h_{j,\sigma}$ take value of 0 or 1 so that $n_{i,\uparrow}n_{i,\downarrow} h_{j,\uparrow}h_{j,\downarrow}$ is 1 if and only if $n_{i,\uparrow}=n_{i,\downarrow}=h_{j,\uparrow}=h_{j,\downarrow}=1$. Therefore, $t \theta_{|i-j|,\sigma} D_i H_j = t \theta_{|i-j|,\sigma} n_{i,\uparrow}n_{i,\downarrow} h_{j,\uparrow}h_{j,\downarrow}$ is equivalent to ReLU$ [t \theta_{|i-j|,\sigma} (n_{i,\uparrow}+n_{i,\downarrow}+h_{j,\uparrow}+h_{j,\downarrow}-3)]$, which is the same as ReLU$ [t \theta_{|i-j|,\sigma} (\sigma_{i}/2+\sigma_{i+N}/2-\sigma_{j}/2-\sigma_{j+N}/2-1)]$. As a result, for each $\eta_{ij,\sigma}$, we associate it with a hidden neuron, such that the weights connecting it to $\sigma_{i},\sigma_{i+N},\sigma_{j},\sigma_{j+N}$ are $t\theta_{|i-j|,\sigma}/2,t\theta_{|i-j|,\sigma}/2,-t\theta_{|i-j|,\sigma}/2,-t\theta_{|i-j|,\sigma}/2$ respectively, the bias is $-t\theta_{|i-j|,\sigma}$ and the activation function is ReLU. In general, for more complicated backflow \cite{Sorella_2008prb_backflow,Sorella_2009prb_backflow,Tocchio_2009jop_backflow} with terms $n_{i,\sigma}h_{i,-\sigma} n_{j,-\sigma}h_{j,\sigma}$, $n_{i,\sigma}n_{i,-\sigma} n_{j,-\sigma}h_{j,\sigma}$ and $n_{i,\sigma}h_{i,-\sigma} h_{j,\sigma}h_{j,-\sigma}$, where $\sigma$ is the spin index, we can use more hidden neurons and represent it in the same way.

After we have the neural network construction for the standard $\eta_{ij,\sigma}$, the term $a^{NN}_{ki} =\sum_{j} \eta_{ij,\sigma}  \phi_{k\sigma}(r_{j,\sigma})$ in Eq.~\eqref{eq:BFNN2} can be realized through an extra layer taking the outputs $\eta_{ij,\sigma}$ to a neuron representing $a_{ki}$ where the weight is given by the single particle orbital values $\phi_{k\sigma}(r_{j,\sigma})$, there is no bias and the activation function is the identity.  This construction shows that the standard backflow parameterization is thus a subset of our three-layer NNB.

\textit{Results.}---We have benchmarked the quality of our NNB on a number of systems including Hubbard models at various sizes and doping (all at $U/t=8$) as well as a frustrated magnet, the Heisenberg model on the Kagome lattice.  
In the main text we focus primarily on the Hubbard model at $n=0.875$ (primarily on the $4 \times 4$ lattice) leaving the additional benchmarks as Supplementary Material  \cite{supp} Sec. IV., \cite{Kagome_ED,Kagome_bkc,Hubbard_benchmark})   
The Hubbard Hamiltonian is
\begin{equation}
H = -t \sum_{i\sigma} (c^{\dagger}_{i\sigma} c_{i+1\sigma} + h.c.) +  \sum_i U n_{i\uparrow} n_{i\downarrow}
\label{eq:Hubbard}
\end{equation}
where we use $U/t=8$.  We compare the results to an optimized unrestricted (i.e. different single particle-orbitals for spin-up and spin-down) Slater Determinant ($\Psi_\textrm{S0}$) as well as a backflow BDG wave function ($\Psi_\textrm{PB}$) which transforms single particle orbitals of each spin by Eq.~\eqref{eq:eta_ij}. The formulation and the variational parameters of each wave function ansatz are summarized in Table.~\ref{tb:ansatz}.  The parameters which aren't optimized, such as the initial set of orbitals $\{\phi_{k\sigma}(r_{i,\sigma})\}$  are obtained for  $\Psi_\textrm{PB}$ and  $\Psi_\textrm{SN}$  by optimizing a restricted Slater-Jastrow wave-function  while $\Psi_\textrm{PN}$ uses orbitals taken from the free hopping Hamiltonian (in practice the nature of the neural net allows for a direct change to the orbitals by altering the bias' on the final layer).

The relative error of the energy of NNB is $1.4\%$ (and $0.66\%$ after variance extrapolation (see Supplementary Material  \cite{supp} Sec. I) which is significantly better then the standard wave-functions (see Fig.~\ref{fig:E_min}(left)).  We examine the effect of the number of hidden neurons $mN$ (see Fig.~\ref{fig:E_min}(right)). We find that at small hidden neuron number, $\Psi_\textrm{PN}$ is much better than $\Psi_\textrm{SN}$ but this advantage eventually largely disappears at large neuron number suggesting that a backflow parameterized with a small neural networks can compensate for the missing pairing in a Slater-determinant.  Surprisingly in the regime we've probed both NNB have energies linear with respect to $1/m$ in spite of the fact that in the $m\rightarrow \infty$ limit, they both must become exact as the FNN could simply put the exact amplitude $\Psi(R)$ on one element of the diagonal \cite{Cybenko_1989MOC_universal}, one on the rest of the diagonal and zero everywhere else. 

\begin{figure}[H]
  \centering
  \includegraphics[scale=0.18]{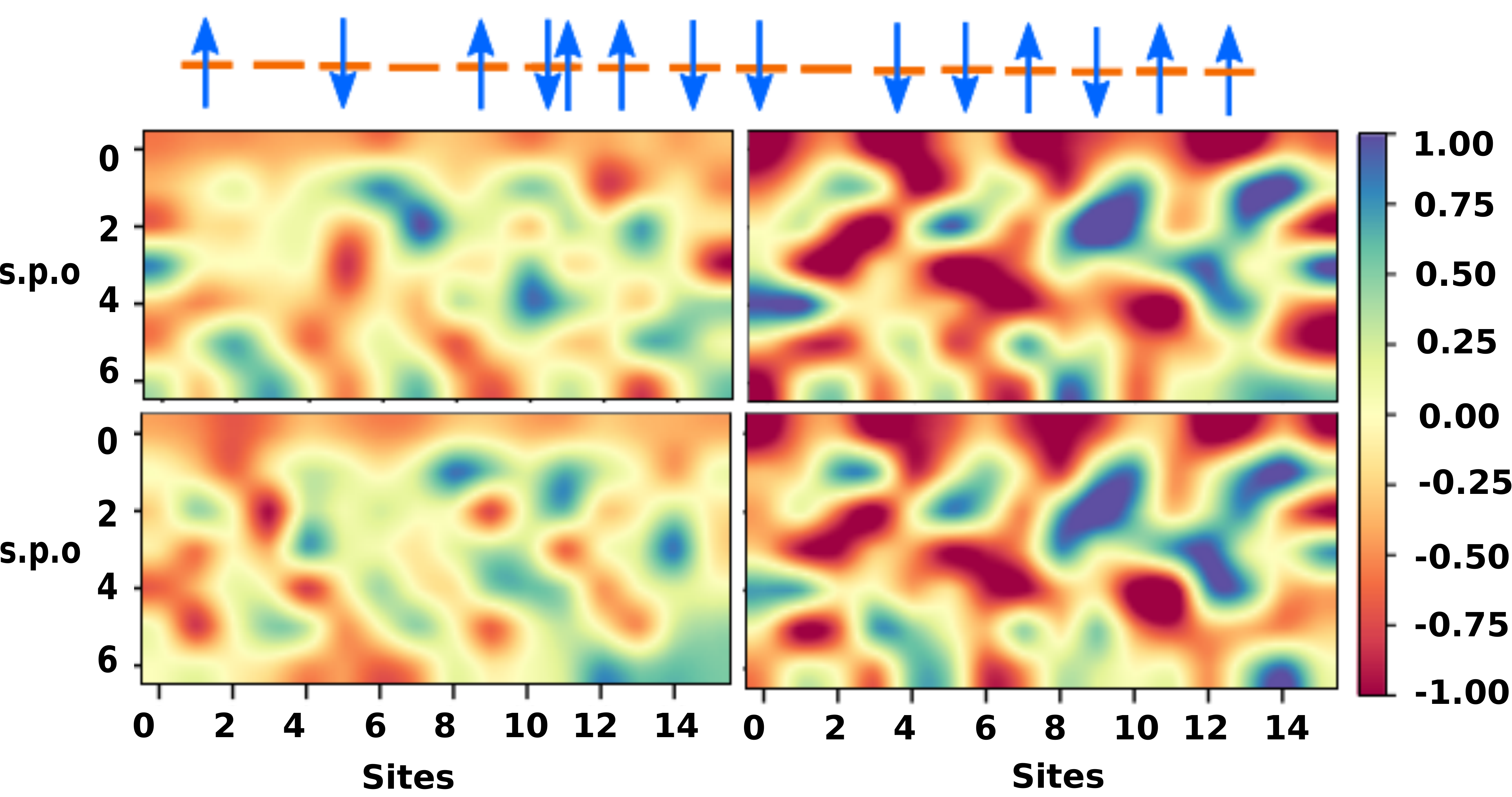}
  \caption{Spin-up (top) and spin-down (bottom)  single particle orbital(s.p.o) for $\Psi_{S0}$ (left) and $\Psi_{SN}$ (right) with 256 hidden neurons on $4 \times 4$ Hubbard model at U/t=8, n=0.875, where the s.p.o are evaluated at the (1d-reshaped) spin-configuration shown.  For the s.p.o, row is orbital index and the column is position index.  
 }\label{fig:BFcorrection_72}
\end{figure}

In addition, we have studied the NNB at $4 \times L$ for $L=\{4,8,12,16\}$ comparing against the DMRG energy (PBC,open) for the $4 \times \infty$ system of $-0.7659 \pm 4 \times 10^{-5}$ per site \cite{Hubbard_stripe}. We find our result very comparable (see Fig.~\ref{fig:E_min} (bottom)) to the DMRG result, especially for system sizes which are commensurate with wave-length 8 stripes\cite{Hubbard_stripe}.

We also investigate how the neural network backflow wave function affects the observables of our system. We see the expectation of double occupancy decreases as $1/m$ (see Supplementary Material  \cite{supp} Sec. I) and the spin and charge densities become significantly more symmetric as the number of hidden neurons increases (see  Fig. ~\ref{fig:Density1}).

To understand the role of neural network in backflow transformation, we investigate how the neural network backflow transformation modify the orbitals. In Fig.~\ref{fig:BFcorrection_72} (right), we notice although the backflow transformation on spin up orbitals and spin down orbitals are performed by two different neural networks, they produce similar backflow transformed orbitals and roughly preserves the symmetry of spin up and spin down for a given configuration. This is different from the optimized  unrestricted Slater Determinant $\Psi_\textrm{S0}$, which breaks the spin up and spin down symmetry significantly (see Fig.~\ref{fig:BFcorrection_72} (Left)).

One feature of using a NNB is the ability to alter the sign-structure of the wave-function.  Here we consider the amount the sign changes between $\Psi_\textrm{SN}$ with 16 hidden neurons and $\Psi_\textrm{S0}$ by evaluating the integral 
\begin{equation}
\frac{\int |\Psi_\textrm{S0}(x)|^2 \text{sgn}(\Psi_\textrm{SN}(x))  \text{sgn}(\Psi_\textrm{S0}(x)) dx}{\int |\Psi_\textrm{S0}(x)|^2 dx }
\end{equation}
which is approximately 0.815 giving a $9\%$ difference between the signs.

\begin{figure}[H]
  \centering
  \includegraphics[scale=0.2]{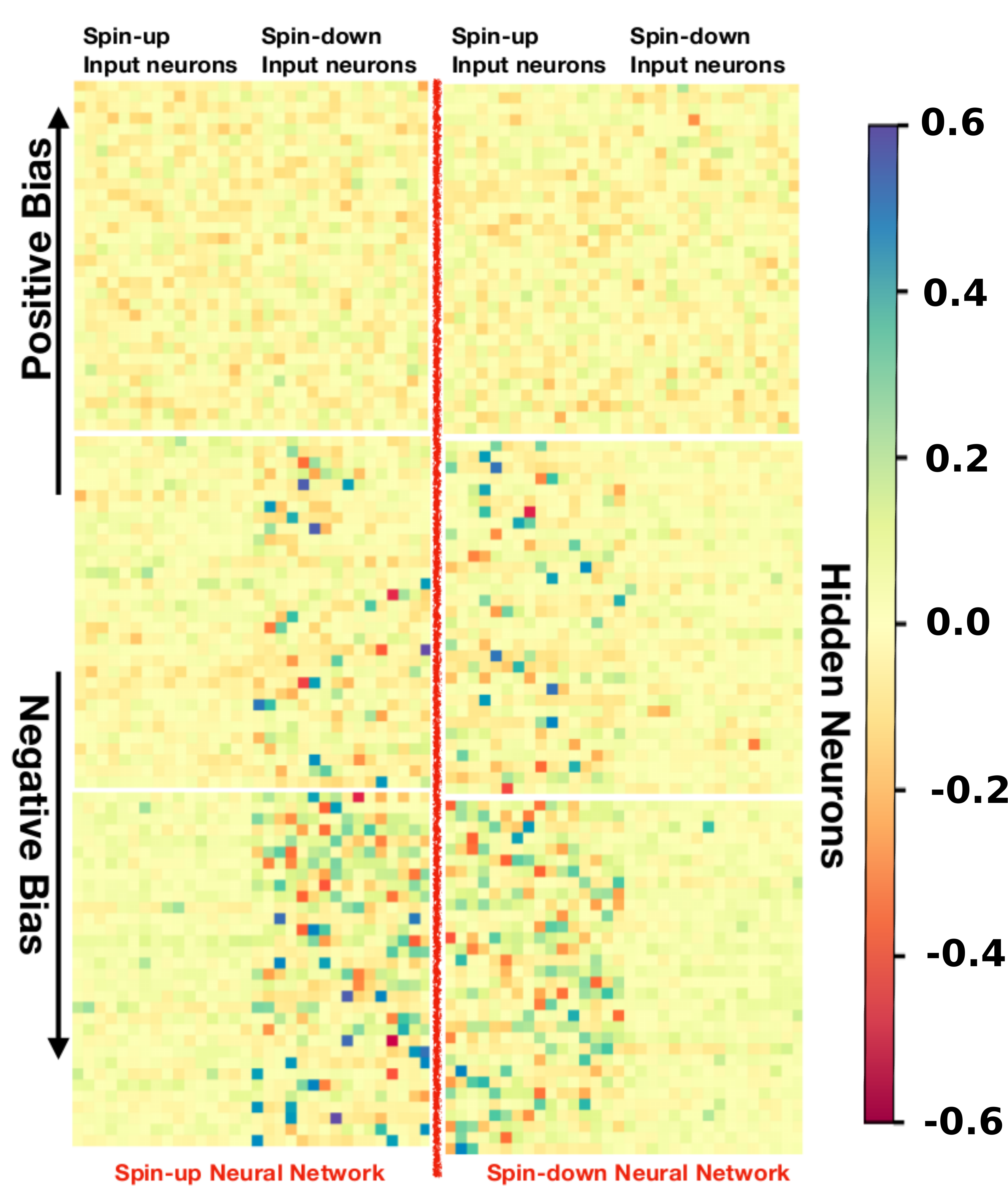}
  \caption{Weights between the input layer and hidden layer for $\Psi_\textrm{SN}$ with 256 hidden neurons for the spin-up (left) and spin-down (right) neural networks on $4 \times 4$ Hubbard model at U/t=8, n=0.875.  Hidden neurons are ordered by their bias and shown are neurons 1-32 (top), 96-128 (middle) and 224-256 (bottom).
  }\label{fig:BF2_b1w1b2w2}
\end{figure}

Furthermore, we open up the $\Psi_\textrm{SN}$ neural network for $m=8$ and analyze the weight between the input layer and the hidden layer, which represents the features that the neural network learns from input. In Fig.~\ref{fig:BF2_b1w1b2w2}, we plot these weights for both the spin-up and spin-down neural networks.  Interestingly the spin-up neural network primarily has large weights connected to the spin-down configurations while the spin-down neural network primarily has large weights connected to the spin-up configurations. This allows the neural network to introduce correlation between spin-up and spin-down configurations. Another observation is that more neurons tend to take large weight in negative bias, and small weight in positive bias.

\textit{Conclusion.}---In this paper, we utilize the generality of artificial neural networks and the physical insight from backflow to develop a new class of wave function ansatz, the neural network backflow wave function, for strongly correlated Fermion systems on lattice. It achieves good performance for Hubbard model at nontrivial filling. We also show improvement on a kagome Heisenberg model in the supplement. While this work has focused on Fermion system on the lattice, the NNB is straightforward to generalize to frustrated spin systems as well as the continuum.  In the latter case, the input could be represented as a lexicographically ordered set of particle locations.  Our work provides a new approach toward combining machine learning methodology with dressed mean-field variational wave-functions which allows us to take simultaneous advantage of their respective strengths.  

\qquad

\textit{Acknowledgement.}---This project is part of the Blue Waters sustained-petascale computing project, which is supported by the National Science Foundation (awards OCI-0725070 and ACI-1238993) and the State of Illinois. Blue Waters is a joint effort of the University of Illinois at Urbana-Champaign and its National Center for Supercomputing Applications. This material is based upon work supported by the U.S. Department of Energy, Office of Science under Award Number FG02-12ER46875. Di acknowledges useful discussion with Ryan Levy, Dmitrii Kochkov, Eli Chertkov, Yusuke Nomura and Giuseppe Carleo. 

\appendix

\section{Supplementary Materials}

\section{Numerical Results on neural network backflow of standard form}\label{app:standard BF}

We have shown that the standard backflow can be represented by a three layer neural network. In practice, we have also investigated the neural network in the following form, which is a direct representation of the standard backflow \cite{Tocchio_2011prb_backflow} 

\begin{equation}
    \phi_{k\sigma}^{b}(r_{i,\sigma};\bm{r}) =  \phi_{k\sigma} + \sum_{j} \eta^{NN}_{ij,\sigma}  \phi_{k\sigma}(r_{j,\sigma}) \label{eq:BFNN1}
\end{equation}
where $\eta^{NN}_{ij,\sigma}$ is represented by a three layer neural network similar to the one used for $a^{NN}_{ij,\sigma}$.

\begin{figure}[H]
  \centering
  \includegraphics[scale=0.75]{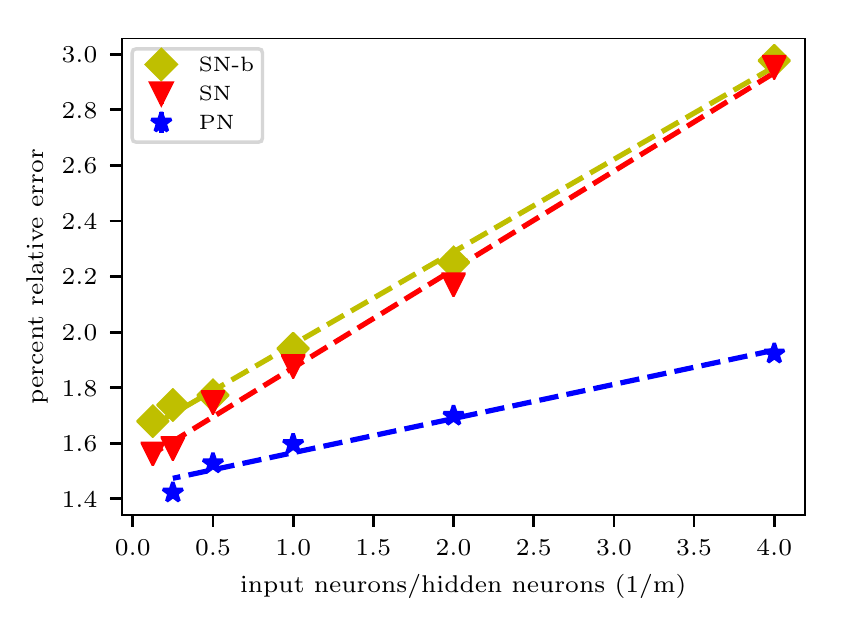}
  \caption{Percentage relative energy error as a function of $1/m$ for $\Psi_\textrm{SN}$, $\Psi_\textrm{SN-b}$ and $\Psi_\textrm{PN}$.  Statistical error bars for energy are shown.
  }\label{fig:N_hidden_neuron2}
\end{figure}

\begin{figure}[H]
  \centering
  \includegraphics[scale=0.75]{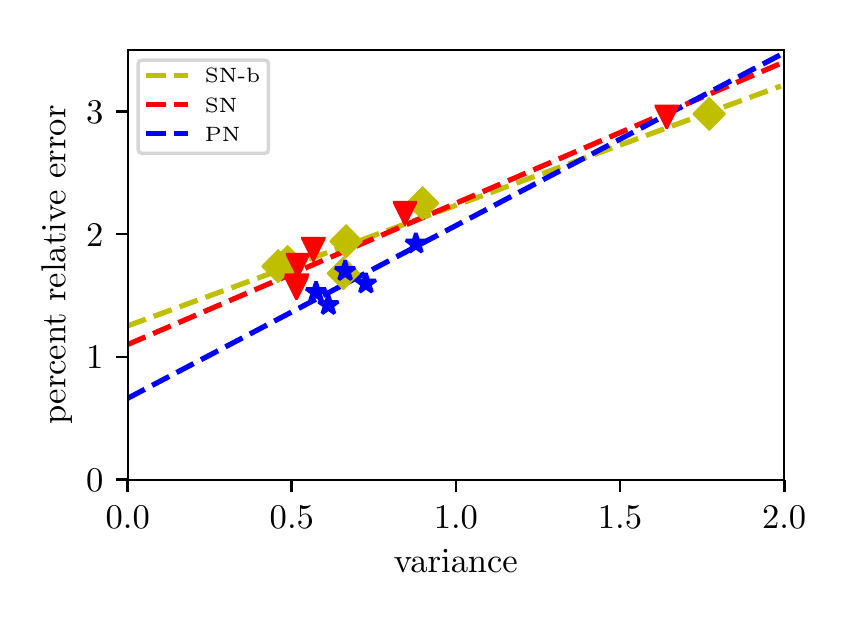}
  \caption{Extrapolation of percentage relative energy error vs. variance of $\Psi_\textrm{SN}$, $\Psi_\textrm{SN-b}$ and $\Psi_\textrm{PN}$.  Statistical error bars for energy and variance are shown. 
  }\label{fig:Energy_var}
\end{figure}

\begin{figure}[H]
  \centering
  \includegraphics[scale=0.75]{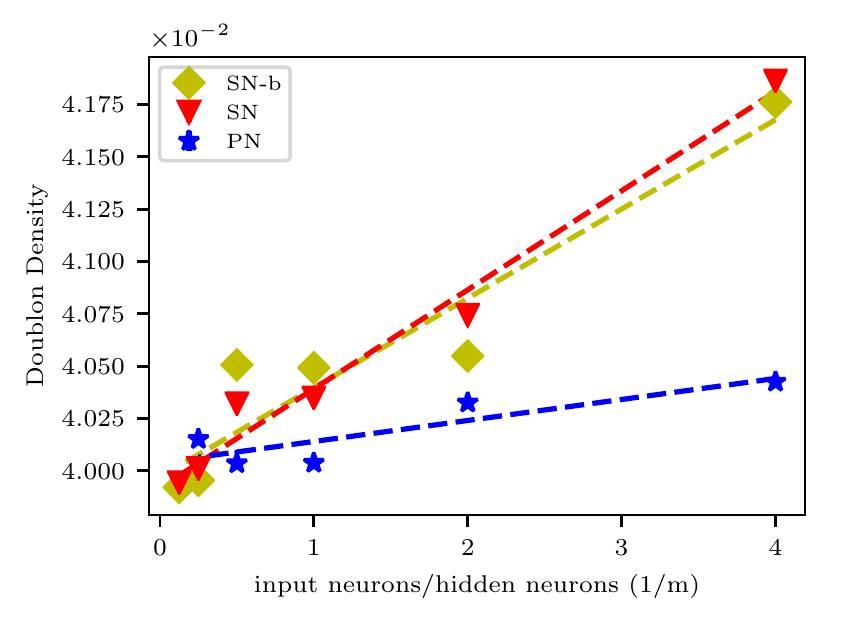}
  \caption{Expectation value of doublon density as a function of $1/m$ for $\Psi_\textrm{SN}$, $\Psi_\textrm{SN-b}$ and $\Psi_\textrm{PN}$. Statistical error bars for doublon density are shown. 
  }\label{fig:Doublon_Average}
\end{figure}

We implement this backflow transformation in the Slater Determinant mean field, which is named as $\Psi_\textrm{SN-b}$. Notice that $\Psi_\textrm{SN-b}$ can also be represented by $\Psi_\textrm{SN}$ since we can define $a^{NN}_{ij,\sigma}=\sum_{j} \eta^{NN}_{ij,\sigma}  \phi_{k\sigma}(r_{j,\sigma})$. In practice, $\Psi_\textrm{SN}$ and $\Psi_\textrm{SN-b}$ have similar performance in computing energy and doublon density (see Fig.~\ref{fig:N_hidden_neuron2}, Fig.~\ref{fig:Energy_var} and Fig.~\ref{fig:Doublon_Average}).

\section{Optimization Scheme}\label{app:optimization}

Given a wave function $\psi$ and Hamiltonian $H$, the energy $E=\frac{\langle \psi|H \psi \rangle }{\langle \psi | \psi \rangle}$. For any variational parameter $t$, to compute the derivative $\frac{\partial E}{\partial t}$ in quantum Monte Carlo, we define the local energy $E_L=\frac{\langle \bm{r}|H \psi \rangle}{\psi(\bm{r})}$, where $\bm{r}$ is a system configuration. Then the derivative can be computed through Monte Carlo sampling.

\begin{equation}
        \frac{\partial E}{\partial t} = 2 \left(\langle E_L (\frac{1}{\psi} \frac{\partial \psi}{\partial t})\rangle -E \langle\frac{1}{\psi} \frac{\partial \psi}{\partial t}\rangle\right)
\end{equation}
where $\langle \hdots \rangle $ is averaged over the Monte Carlo samples.

To update the parameter $t$ in the $k+1$ -th iteration, we use the sign of the gradient with a random magnitude as follows.

\begin{equation}
	t_{k+1}= t_k - \alpha \gamma \frac{|\partial E/ \partial t_k|}{\partial E/ \partial t_k}
\end{equation}
where $\alpha$ is a random number in the range $(0,1)$, $\gamma$ is the step size, $t_k$ and $t_{k+1}$ are the values for variational parameter $t$ at the $k$-th and $k+1$-th iteration.  The scheme that alpha is random is inspired by the method described in \cite{sandvik_2007_SR}.  This method can help to get the optimization out of shallow local minima and is less sensitive to noise then standard gradient descent approaches. As a practical matter, we have optimized using this approach, gradient descent and stochastic reconfiguration and find that this approach is the best tradeoff between speed and quality of optimization. In practice, we choose $\gamma=0.001$, weights and bias' of the neural network are initialized from uniform distribution in the range of $(-0.005,0.005)$.

Notice that to represent Slater Determinant with NNB, one can set the bias of the output layer in NNB to be the orbitals while the other parameters to be zero. This observation can reduce the difficulty of optimization by initializing NNB with the optimized Slater Determinant, where the bias of the output layer is the optimized orbitals and the other parameters are small random numbers around zero. This trick is helpful for optimization in large systems.

\section{Backward propagation of neural network backflow}\label{app:backprop}

From the optimization scheme, it is clear that we need to compute $\frac{1}{\psi} \frac{\partial \psi}{\partial t}$. In this section, we give the details on how to compute this quantity by backward propagation.

In neural network algorithm, forward propagation refers to the evaluation of the network while backward propagation is essentially performing chain rule to compute derivatives. For forward propagation, the values in the $i+1$-th layer are computed from the values in the $i$-th layer under matrix operation and activation function, so that the output layer's values can be obtained from the input layer's values. For backward propagation, intuitively one can understand it as backward propagation of 'errors', where the 'errors' in the $i$-th layer are computed from the 'errors' in the $i+1$-th layer. More accurately, the 'errors', such as ${}^i\delta^{l}_p$ in Eq.~\eqref{eq:backward_prop}, are related to the derivatives of the parameters. It indicates that the derivatives  of parameters in the $i$-th layer level can be computed from the derivatives of in the $i+1$-th layer level, which is essentially the chain rule of computing derivatives. A detailed description of applying backward propagation in NNB is given as follows.

Consider backward propagation in a neural network with N layers. Denote the value in each layer by a set of vectors ${a^0, a^1, ..., a^{N-1}}$. Denote the value before activation in each layer (counting from the 2nd layer) by a set of vectors ${z^1,z^2,...,z^{N-1}}$. Denote the activation function in each layer (counting from the 2nd layer) by a set of functions ${f_1,f_2,...,f_{N-1}}$. Denote the weights by a set of weights (counting from the 2nd layer)  by a set of matrix ${w^1,w^2,...,w^{N-1}}$ and the bias by a set of vectors by a set of vector ${b^1,b^2,...,b^{N-1}}$. Then $a^0$ is the input configuration, for $1 < l \leq N-1$ , forward propagation reads

\begin{align}
        z^{l} &= w^{l} a^{l-1} + b^{l} \\
        a^{l} &= f_{l} (z^{l}) 
\end{align}

To calculate derivative, we use backward propagation. For the i-th output in the final layer, denote the error of each layer (counting from the 2nd layer) with respect to this output by a set of vectors ${{}^i\delta^1,{}^i\delta^2,...,{}^i\delta^{N-1}}$. Set ${}^i\delta^{N-1}= f'_{N-1}(z^{N-1}_i)$. For $1 \leq l \leq N-1$, backward propagation reads

\begin{align}
        {}^i\delta^{l-1} &= ((w^{l})^T  {}^i\delta^{l} ) * f'_{l-1}(z^{l-1}) \\
        \frac{\partial a^{N-1}_i}{\partial b^{l}_p} &= {}^i\delta^{l}_p \\\label{eq:backward_prop}
        \frac{\partial a^{N-1}_i}{\partial w^{l}_{pq}} &= a^{l-1}_q \otimes {}^i\delta^{l}_p
\end{align}
where $*$ is the element-wise multiplication and the last equation uses outer product $\otimes$.

To transfer backward propagation to $\frac{1}{\psi} \frac{\partial \psi}{\partial t}$, we should view it as a special mapping on top of $a^{N-1}$. Therefore, we only need to choose a different ${}^i\delta^{N-1}$ and the rest will be the same as above. First, we consider the Slater-Determinant type neural network backflow.

\begin{equation}
\begin{split}
\frac{1}{\psi_{SD}} \frac{\partial \psi_{SD}(\bm{r})}{\partial t} & =   \frac{1}{det[M^{SD,\uparrow}]} \frac{\partial \det[M^{SD,\uparrow}]}{\partial t} \\
&+\frac{1}{det[M^{SD,\downarrow}]} \frac{\partial \det[M^{SD,\downarrow}]}{\partial t}; \\
\end{split}
\end{equation}
where $M_{ik}^{SD,\sigma} =\phi^{b}_{k\sigma}(r_{i\sigma})$, $t$ is any variational parameter $\{w^l_{pq},b^l_p\}$. Under the Einstein notation, the chain rule gives rise to, 

\begin{align}
& \frac{1}{det[M^{SD,\sigma}]} \frac{\partial \det[M^{SD,\sigma}]}{\partial t} \\
& =\frac{1}{det[M^{SD,\sigma}]} \frac{\partial \det[M^{SD,\sigma}]}{\partial \phi^{b}_{k\sigma}(r_{i\sigma})} \frac{\partial \phi^{b}_{k\sigma}(r_{i\sigma})}{\partial t} \\
& =tr( (M^{SD,\sigma})^{-1} \frac{\partial \det[M^{SD,\sigma}]}{\partial \phi^{b}_{k\sigma}(r_{i\sigma})} ) \frac{\partial \phi^{b}_{k\sigma}(r_{i\sigma})}{\partial t} \\
& = (M^{SD,\sigma})^{-1}_{ki} \frac{\partial \phi^{b}_{k\sigma}(r_{i\sigma})}{\partial t}
\end{align}

For $\Psi_\textrm{SN}$, we have
\begin{align}
& \frac{1}{det[M^{SD,\sigma}]} \frac{\partial \det[M^{SD,\sigma}]}{\partial t} \\
& = (M^{SD,\sigma})^{-1}_{ki} \frac{\partial \phi^{b}_{k\sigma}(r_{i\sigma})}{\partial a^{NN}_{ki,\sigma}(\bm{r})} \frac{a^{NN}_{ki,\sigma}(\bm{r})}{\partial t} \\
& = (M^{SD,\sigma})^{-1}_{ki}  \frac{a^{NN}_{ki,\sigma}(\bm{r})}{\partial t}
\end{align}
Therefore, we set ${}^i\delta^{N-1}_{ki}=(M^{SD,\sigma})^{-1}_{ki} f'_{N-1}(z^{N-1}_{ki})$ and perform the backward propagation.Notice that we use a matrix index $ki$ for simplicity here. In practice, a matrix will be reshaped into an array for computation.

For $\Psi_\textrm{SN-b}$, we have
\begin{align}
& \frac{1}{det[M^{SD,\sigma}]} \frac{\partial \det[M^{SD,\sigma}]}{\partial t} \\
& = (M^{SD,\sigma})^{-1}_{ki} \frac{\partial \phi^{b}_{k\sigma}(r_{i\sigma})}{\partial \eta^{NN}_{ij,\sigma}(\bm{r})} \frac{\eta^{NN}_{ij,\sigma}(\bm{r})}{\partial t}
\end{align}
We then set ${}^i\delta^{N-1}_{ij}=(M^{SD,\sigma})^{-1}_{ki} \frac{\partial \phi^{b}_{k\sigma}(r_{i\sigma})}{\partial \eta^{NN}_{ij,\sigma}(\bm{r})}  f'_{N-1}(z^{N-1}_{ij})$ and perform the backward propagation. Notice that we use a matrix index $ij$ for simplicity here. In practice, a matrix will be reshaped into an array for computation.

Next, we consider Bogoliubov de Gennes type neural network backflow wave-functions $\Psi_\textrm{PN}$, 

\begin{equation}
\frac{1}{\psi_{BDG}} \frac{\partial \psi_{BDG}(\bm{r})}{\partial t}  =   \frac{1}{det[\Phi]} \frac{\partial \det[\Phi]}{\partial t}
\end{equation}
where $\Phi_{ij} = \sum^{N}_{k,l=1} \phi^b_{k\uparrow}(r_{i,\uparrow}) S_{kl}  \phi^b_{l\downarrow}(r_{j,\downarrow})$, $t$ is any variational parameter $\{w^l_{pq},b^l_p\}$ of $a^{NN}_{ij,\sigma}$ and $d^{NN}_{kl}$.

For the case that $t$ is a parameter of $a^{NN}_{ki,\uparrow}$, we rewrite $\Phi_{ij} = \sum^{N}_{k,l=1} \phi^b_{k\uparrow}(r_{i,\uparrow}) R_{kj}$, where $R_{kj}= \sum^{N}_l S^{NN}_{kl}  \phi^b_{l\downarrow}(r_{j,\downarrow})$, $\phi^b_{k\uparrow}(r_{i,\uparrow})$ is given by Eq.~(5) in the main body of the paper and $S^{NN}_{kl}=S_{kl}+d^{NN}_{kl}$. The chain rule gives rise to

\begin{align}
&\frac{1}{det[\Phi]} \frac{\partial \det[\Phi]}{\partial t}\\
&=\frac{1}{det[\Phi]} \frac{\partial \det[\Phi]}{\partial \phi^b_{k\uparrow}(r_{i,\uparrow})} \frac{\partial \phi^b_{k\uparrow}(r_{i,\uparrow})}{\partial a_{kl}} \frac{\partial a_{kl}}{\partial t} \\
&= \frac{1}{det[\Phi]} \frac{\partial det[\Phi]}{\partial \phi^b_{k\uparrow}(r_{i,\uparrow})} \frac{\partial a_{kl}}{\partial t} \\
&= \Phi^{-1}_{ji} R_{kj} \frac{\partial a_{kl}}{\partial t} 
\end{align}
We then set ${}^i\delta^{N-1}_{ki}=\Phi^{-1}_{ji} R_{kj}  f'_{N-1}(z^{N-1}_{ki})$ and perform the backward propagation.
For $t$ is a parameter of $a^{NN}_{ki,\downarrow}$, similarly we define $Q_{il}= \sum^{N}_k (\phi^b_{k\downarrow})^T(r_{i,\downarrow}) S^{NN}_{kl}$ and then set ${}^i\delta^{N-1}_{ki}=\Phi^{-1}_{ij} Q_{jk}  f'_{N-1}(z^{N-1}_{ki})$.

For $t$ is a parameter of $d^{NN}_{kl}$,

\begin{align}
&\frac{1}{det[\Phi]} \frac{\partial \det[\Phi]}{\partial t}\\
&=\frac{1}{det[\Phi]} \frac{\partial \det[\Phi]}{\partial S^{NN}_{kl}} \frac{\partial S^{NN}_{kl}}{\partial d_{kl}} \frac{\partial d_{kl}}{\partial t} \\
&= \frac{1}{det[\Phi]} \frac{\partial \det[\Phi]}{\partial S^{NN}_{kl} } \frac{\partial d_{kl}}{\partial t} \\
&=  \phi^b_{k\uparrow}(r_{i,\uparrow}) (\Phi^{-1})^T_{ij} \phi^b_{l\downarrow}(r_{j,\downarrow})  \frac{\partial d_{kl}}{\partial t} 
\end{align}
We set ${}^i\delta^{N-1}_{ki}=\phi^b_{k\uparrow}(r_{i,\uparrow}) (\Phi^{-1})^T_{ij} \phi^b_{l\downarrow}(r_{j,\downarrow}) f'_{N-1}(z^{N-1}_{ki})$ and perform the backward propagation.

\section{Neural Network Backflow on various systems}

\begin{figure}[H]
  \centering
  \includegraphics[scale=0.49]{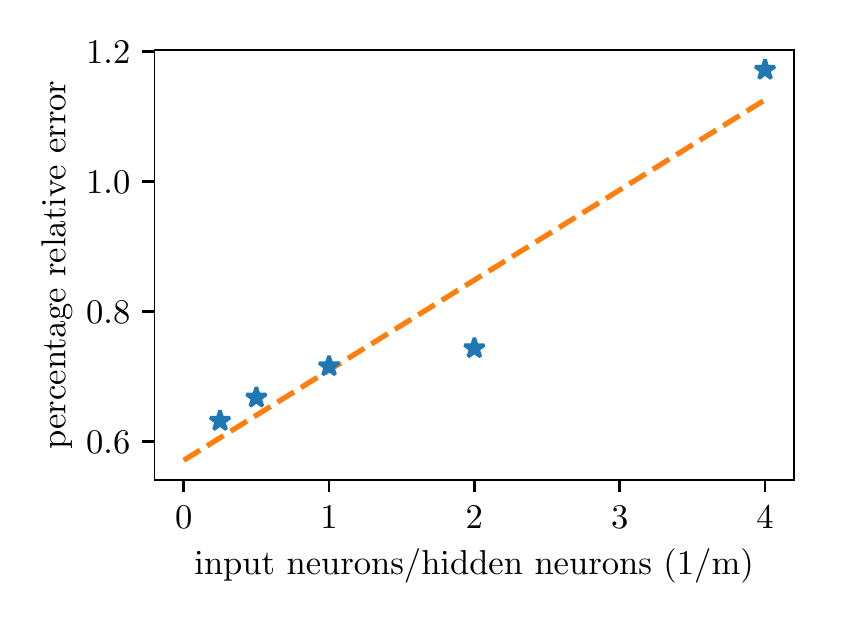}
  \includegraphics[scale=0.49]{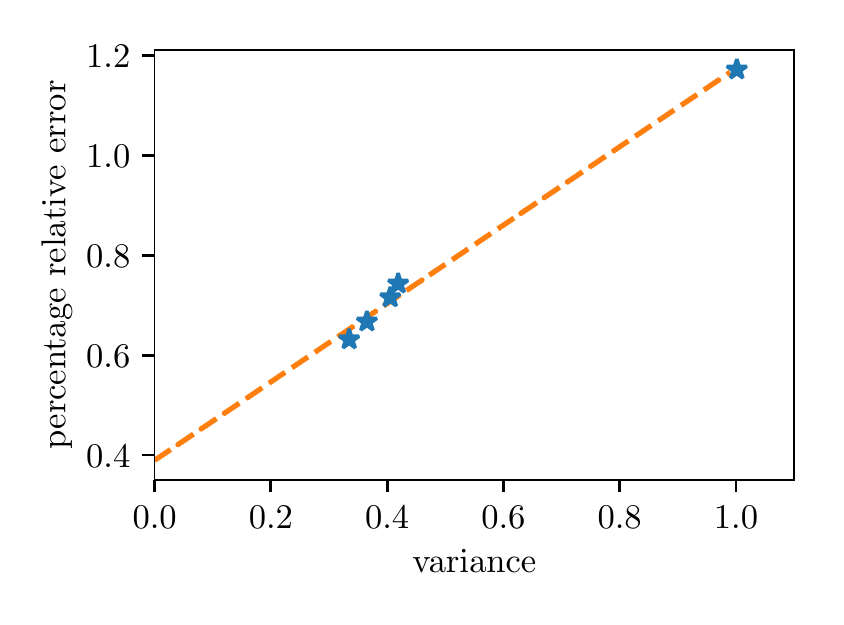}
  \caption{$\Psi_\textrm{SN}$ on $4 \times 4$ Hubbard model with filling $n=0.75$. \textbf{Left}: Percentage relative energy error as a function of $1/m$. \textbf{Right}: Percentage relative energy error as a function of variance. Statistical error bars for energy are shown.
  }\label{fig:Energy_4x4_vmc}
\end{figure}

\begin{figure}[H]
  \centering
  \includegraphics[scale=0.49]{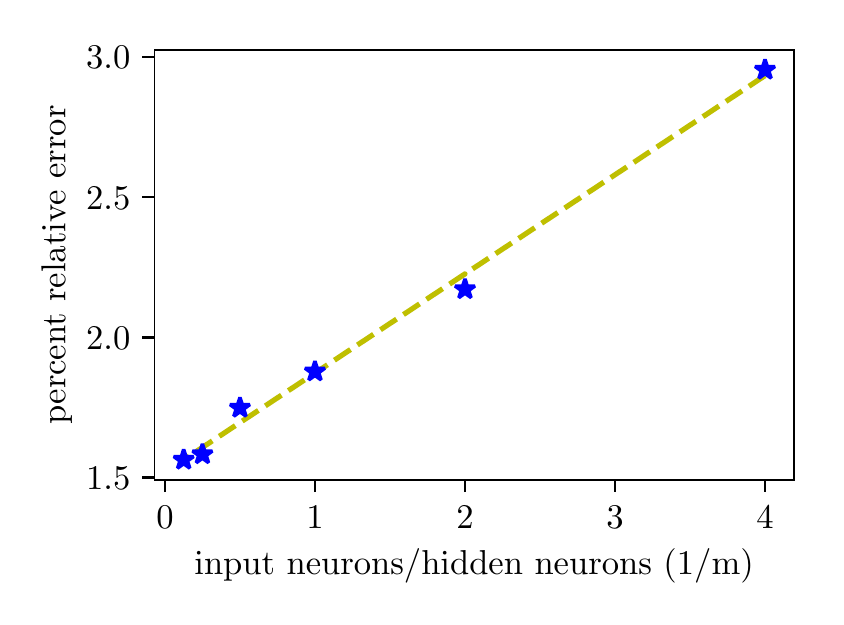}
  \includegraphics[scale=0.49]{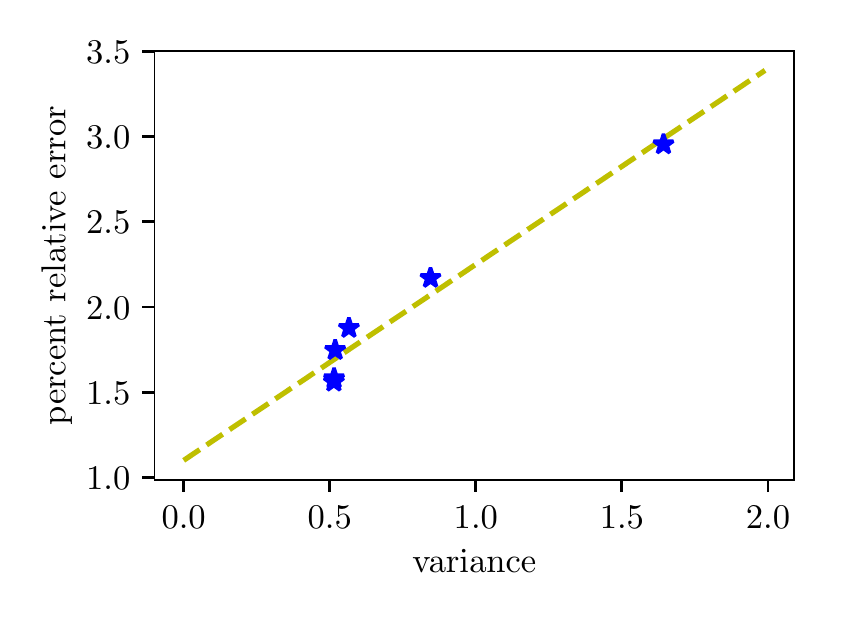}
  \caption{$\Psi_\textrm{SN}$ on $4 \times 4$ Hubbard model with filling $n=0.875$. \textbf{Left}: Percentage relative energy error as a function of $1/m$. \textbf{Right}: Percentage relative energy error as a function of variance. Statistical error bars for energy are shown.
  }\label{fig:Energy_4x4_n0875_vmc}
\end{figure}

\begin{figure}[H]
  \centering
  \includegraphics[scale=0.49]{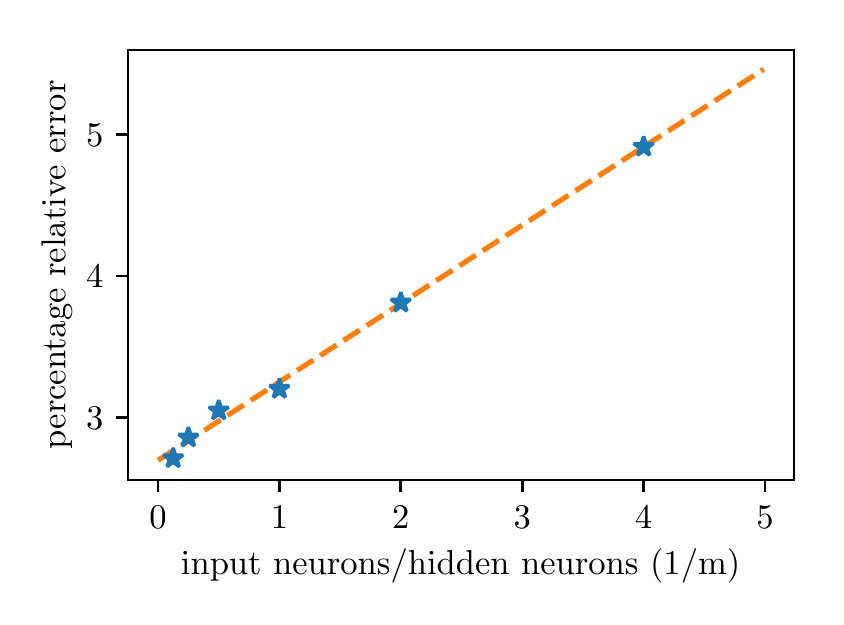}
  \includegraphics[scale=0.49]{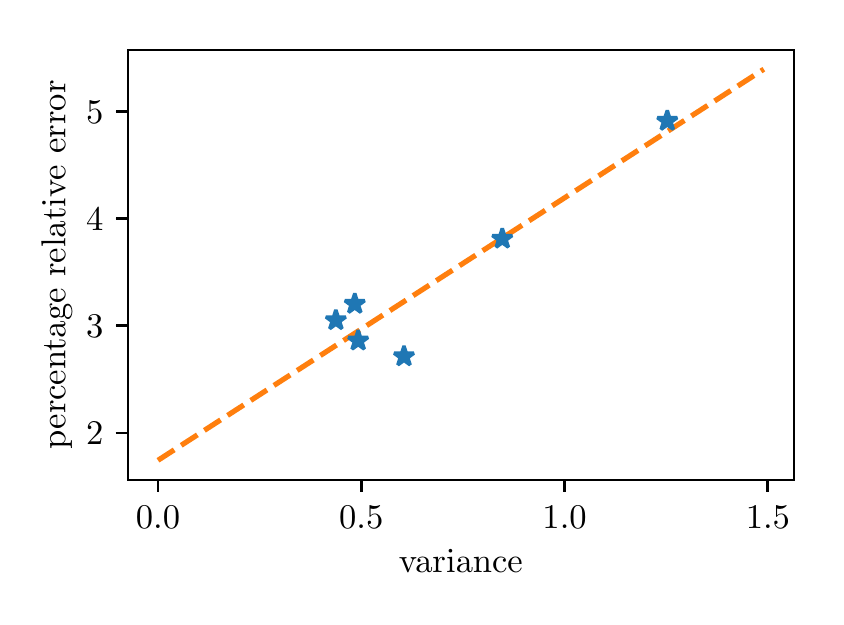}
  \caption{$\Psi_\textrm{SN}$ on $4 \times 4$ Hubbard model with filling $n=1.0$. \textbf{Left}: Percentage relative energy error as a function of $1/m$. \textbf{Right}: Percentage relative energy error as a function of variance. Statistical error bars for energy are shown.
  }\label{fig:Energy_4x4_n1_vmc}
\end{figure}

\begin{figure}[H]
  \centering
  \includegraphics[scale=0.49]{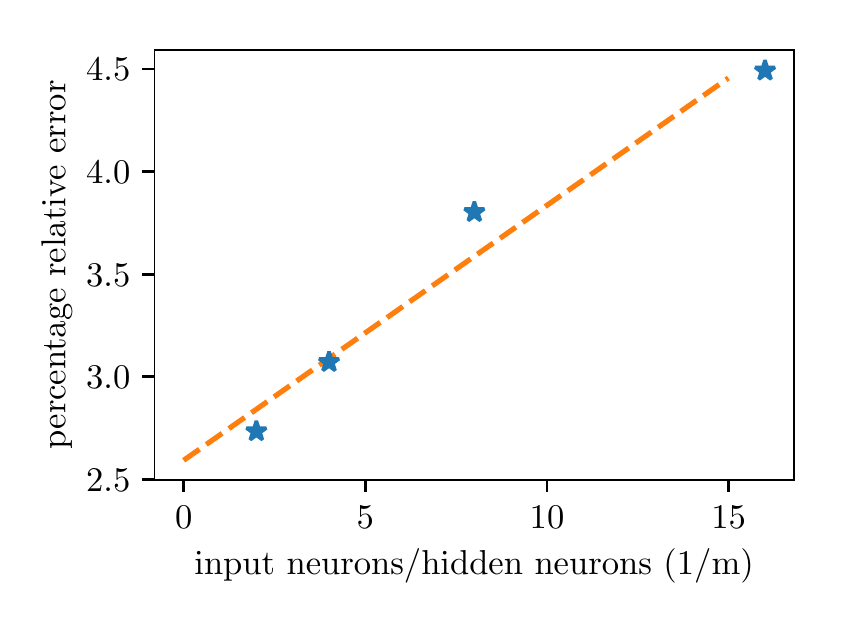}
  \includegraphics[scale=0.49]{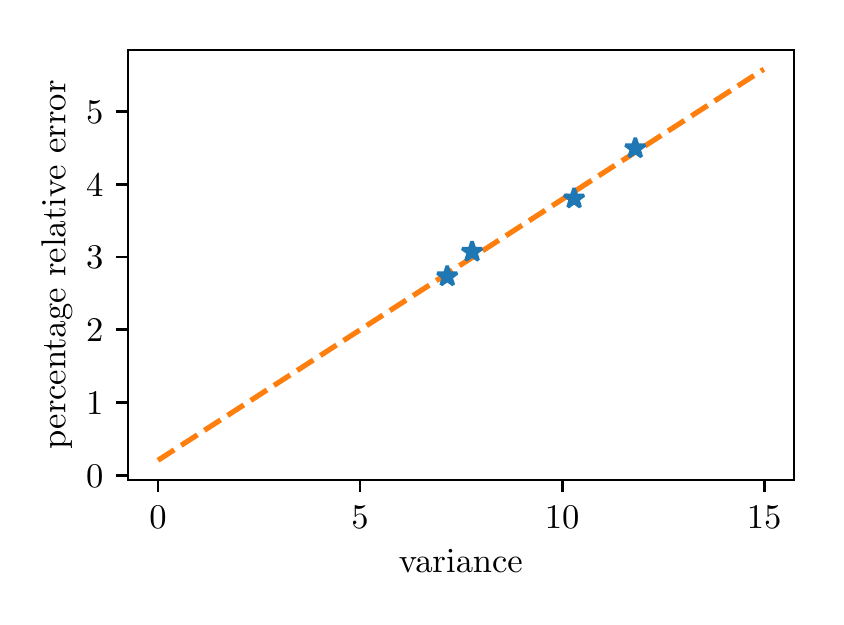}
  \caption{$\Psi_\textrm{SN}$ on $16 \times 4$ Hubbard model with filling $n=0.875$. \textbf{Left}: Percentage relative energy error as a function of $1/m$. \textbf{Right}: Percentage relative energy error as a function of variance. Statistical error bars for energy are shown.
  }\label{fig:Energy_16x4_vmc}
\end{figure}

\begin{figure}[H]
  \centering
  \includegraphics[scale=0.49]{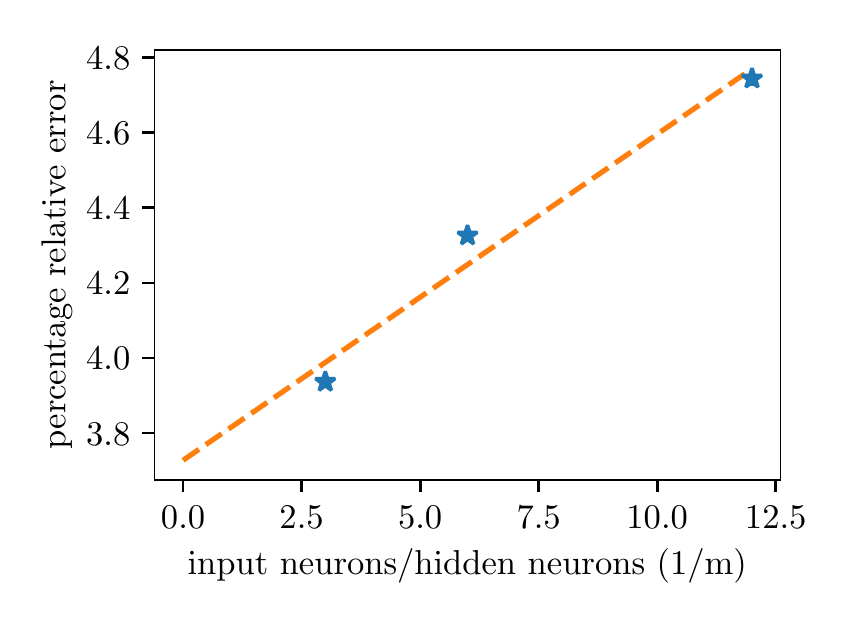}
  \includegraphics[scale=0.49]{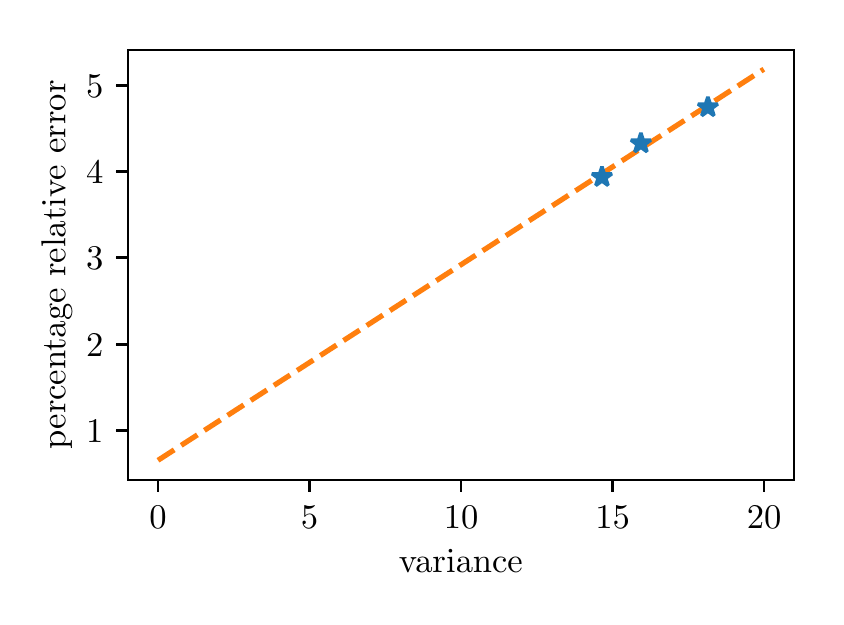}
  \caption{$\Psi_\textrm{SN}$ on $12 \times 8$ Hubbard model with filling $n=0.875$. \textbf{Left}: Percentage relative energy error as a function of $1/m$. \textbf{Right}: Percentage relative energy error as a function of variance. Statistical error bars for energy are shown.
  }\label{fig:Energy_12x8_n0875_vmc}
\end{figure}

\begin{figure}[H]
  \centering
  \includegraphics[scale=0.49]{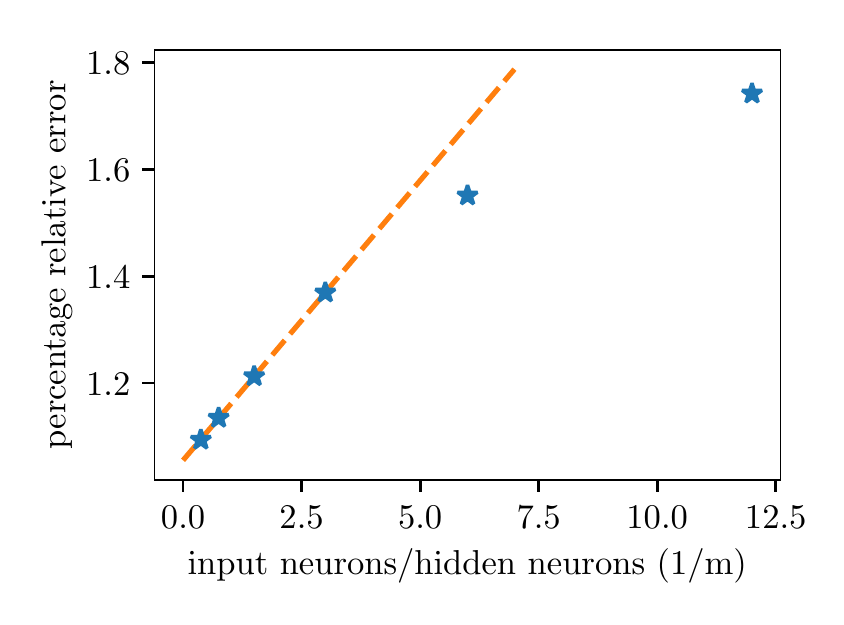}
  \includegraphics[scale=0.49]{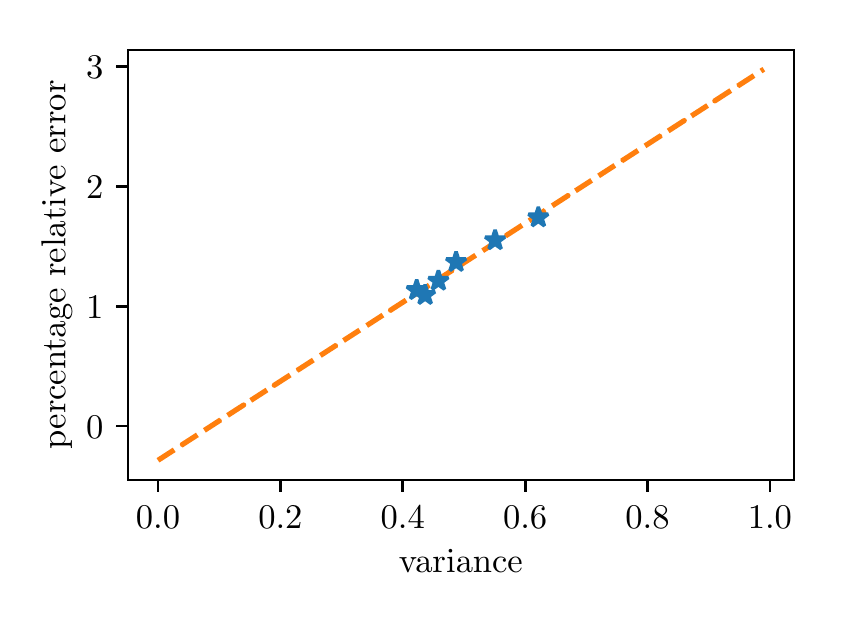}
  \caption{$\Psi_\textrm{SN}$ on $4 \times 4 \times 3$ Kagome model. \textbf{Left}: Percentage relative energy error as a function of $1/m$. \textbf{Right}: Percentage relative energy error as a function of variance. Statistical error bars for energy are shown.
  }\label{fig:Energy_Kagome}
\end{figure}

\begin{figure}[H]
  \centering
  \includegraphics[scale=0.49]{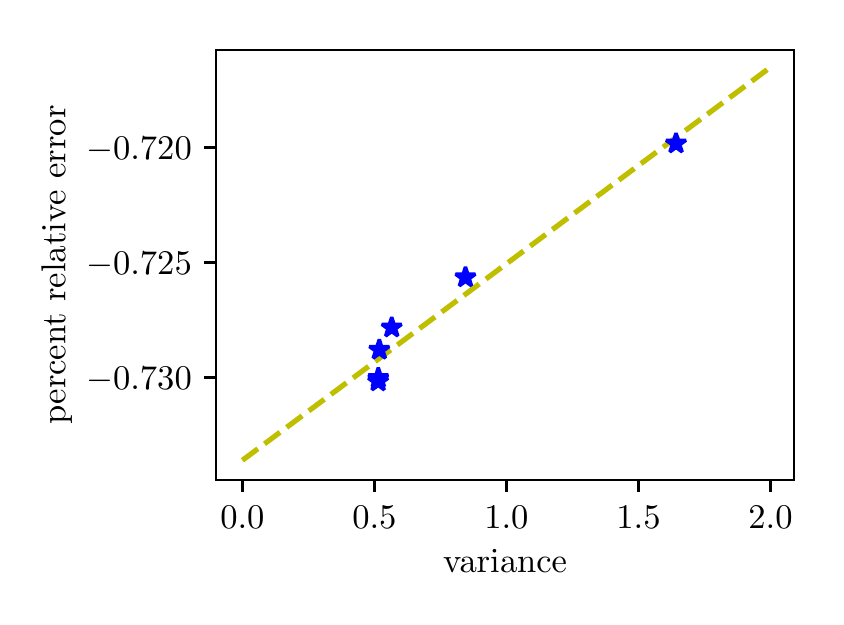}
  \includegraphics[scale=0.49]{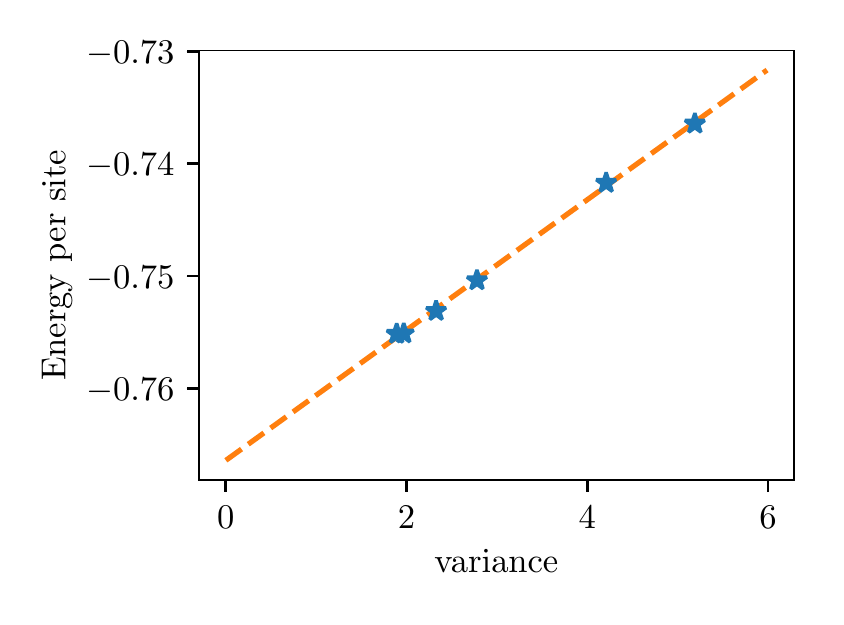}
  \caption{$\Psi_\textrm{SN}$ energy per site vs. variance on Hubbard model at U/t=8, n=0.875. \textbf{Left}: $4 \times 4 $ Hubbard. \textbf{Right}: $8 \times 4 s$ Hubbard. Statistical error bars for energy are shown.
  }\label{fig:size4_size8}
\end{figure}

\begin{figure}[H]
  \centering
  \includegraphics[scale=0.49]{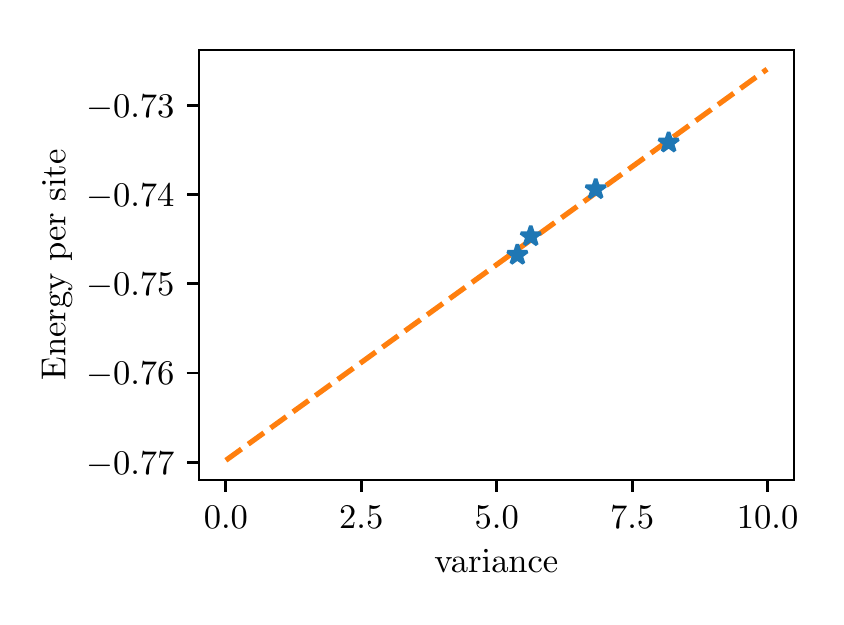}
  \includegraphics[scale=0.49]{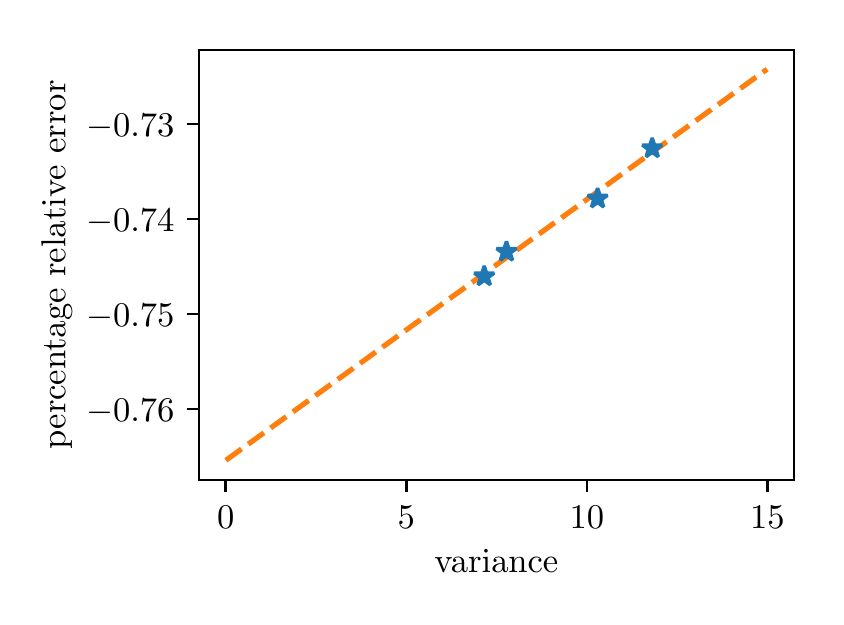}
  \caption{$\Psi_\textrm{SN}$ energy per site vs. variance on Hubbard model at U/t=8, n=0.875. \textbf{Left}: $12 \times 4 $ Hubbard. \textbf{Right}: $16 \times 4 $ Hubbard. Statistical error bars for energy are shown.
  }\label{fig:size12_size16}
\end{figure}

To further test the ability of NNB, we have optimized $\Psi_\textrm{SN}$ in various systems: (i) $4 \times 4$ Hubbard model at $U/t=8$ with $n=0.75$ and $n=1.0$ (See Fig.~\ref{fig:Energy_4x4_vmc},~\ref{fig:Energy_4x4_n0875_vmc},~\ref{fig:Energy_4x4_n1_vmc}). (ii) $8 \times 4$, $12 \times 4$, $16 \times 4$ Hubbard model with $n=0.875$ and $U/t=8$ (See Fig.~\ref{fig:size4_size8}, ~\ref{fig:size12_size16},~\ref{fig:Energy_16x4_vmc}). (iii) $12 \times 8$ Hubbard model at $U/t=8$ with $n=0.875$ (See Fig.~\ref{fig:Energy_12x8_n0875_vmc}). (iv) $4 \times 4 \times 3$ Kagome Heisenberg model with $J=1$ ((See Fig.~\ref{fig:Energy_Kagome})). The optimization has been implemented with the RMSPROP \cite{opt} method.

For the $4 \times 4$ Hubbard model, we find that the smallest doping ($n=0.75$) does best with an error (variance extrapolated error) of $0.631\%$ ($0.233\%$) with the largest doping ($n=1.0$) doing the worst giving a $2.714\%$ ($1.745\%$) error (See Fig.~\ref{fig:Energy_4x4_vmc},~\ref{fig:Energy_4x4_n0875_vmc},~\ref{fig:Energy_4x4_n1_vmc} for details). The results are compared with the exact diagonalization data \cite{Dagotto_1992prb_ED}.

We consider both the $16\times 4$ and $12 \times 8$ lattice at $U=8$ and $n=0.875$.  Because there are no exact answers at these system sizes, we compare against approximate (not necessarily variationally upper-bounded) AFQMC.  For our $16 \times 4$ system, using a number of hidden neurons  $n_h=\{8,16,32,64\}$, we find the energy decreases from  -46.211 (for the optimized Slater determinant) to -47.745 ($n_h=64$). The variance extrapolated energy is approximately -49 which is reasonably close to the twist-averaged boundary condition AFQMC result of -49.088\cite{Hubbard_benchmark} (See Fig.~\ref{fig:Energy_16x4_vmc}). In the $12\times 8$ system we use $n_h=\{8,16,32\}$ with relative errors against the same system in AFQMC \cite{Hubbard_stripe} of $6.3\%$ for the optimized Slater determinant, $3.94\%$ for the $n_h=32$ NNB and $0.655\%$ after variance extrapolation  (See Fig.~\ref{fig:Energy_12x8_n0875_vmc}).  NNB significantly improves on the standard Slater-Jastrow methodology and the variance extrapolation result is competitive with  other techniques.

Finding wave-functions for the ground state of the Heisenberg model on the Kagome lattice is challenging. The best per site energy of a  projected BDG result on a $4\times 4\times 3$ lattice is approximately -0.4305 \cite{Kagome_bkc} compared to  the exact answer of -0.4387 \cite{Kagome_ED}. We have implemented the neural network backflow on top of the Slater Determinant with  $n_h = \{8,16,32,64,128,256\}$. The associated energies ranges from $-0.4311$ ($n_h=8$) to $-0.4339$ ($n_h=256$), giving an improvement of the relative error of approximately 50$\%$  (See Fig.~\ref{fig:Energy_Kagome}).

The results are summarized in Table \ref{tb:energy} and \ref{tb:energy_fs}.

\begin{table*}[t]
  \centering
  \scalebox{0.9}{
  \begin{tabular}{|c|c|c|c|c|} \hline
  Relative energy error & Slater-Jastrow & NNB  & NNB neuron extrapolation & NNB variance extrapolation \\\hline
  $4 \times 4$ Hubbard, n=0.75 & $(3.6 \pm 6 \times 10^{-4}) \%$ & $(0.631 \pm 8 \times 10^{-4}) \%$  & 0.607$\%$ & 0.233$\%$ \\\hline
  $4 \times 4$ Hubbard, n=0.875 & $(5.4 \pm 2 \times 10^{-3}) \%$ & $(1.156 \pm 2 \times 10^{-3}) \%$  & 1.152$\%$ & 1.101$\%$ \\\hline
  $4 \times 4$ Hubbard, n=1.0 & $(5.9 \pm 2 \times 10^{-3}) \%$ &
$(2.714 \pm 5 \times 10^{-3}) \%$  & 2.700$\%$ & 1.745$\%$ \\\hline
$16 \times 4$ Hubbard, n=0.875 & $(5.9 \pm 2 \times 10^{-3}) \%$ & $(2.734 \pm 8 \times 10^{-3}) \%$  & 2.592$\%$ & 0.209$\%$ \\\hline
$12 \times 8$ Hubbard, n=0.875 & $(6.3 \pm 3 \times 10^{-3}) \%$ & $(3.94 \pm 10^{-2}) \%$  & 3.727$\%$ & 0.655$\%$ \\\hline
$4 \times 4 \times 3$ Kagome & $(1.8 \pm 10^{-5}) \%$ & 
$(1.093 \pm 4 \times 10^{-3}) \%$  & 1.055$\%$ & 0.286$\%$ \\\hline
  \end{tabular}}
  \caption{Neural Network Backflow performance on various systems}
  \label{tb:energy}
\end{table*}

\begin{table*}[t]
  \centering
  \begin{tabular}{|c|c|c|c|c|} \hline
  Hubbard model energy per site & Slater-Jastrow & NNB  & NNB variance extrapolation \\\hline
  $4 \times 4$, U/t=8, n=0.875 & $-0.702 \pm 10^{-5}$  & $-0.730 \pm 8 \times 10^{-6} $  & -0.734 \\\hline
  $8 \times 4$, U/t=8, n=0.875 & $-0.719 \pm 10^{-5}$  & $-0.755 \pm 4 \times 10^{-5} $  & -0.767 \\\hline
  $12 \times 4$, U/t=8, n=0.875 & $-0.722 \pm 2 \times 10^{-5}$  & $-0.746 \pm 5 \times 10^{-5} $  & -0.770 \\\hline
  $16 \times 4$, U/t=8, n=0.875 & $-0.722 \pm 10^{-5}$  & $-0.746 \pm 6 \times 10^{-5}$  & -0.765
 \\\hline
  \end{tabular}
  \caption{Finite size effect study of Neural Network Backflow}
  \label{tb:energy_fs}
\end{table*}

\bibliography{BackflowNN}

\end{document}